\documentclass[aps,prc,twocolumn,showpacs,amsmath,amssymb,nofootinbib]{revtex4}
\usepackage{epsfig,colordvi}

\usepackage{graphicx}
\usepackage{dcolumn}
\usepackage{bm}
\usepackage{epsfig}
\usepackage{array}
\usepackage{amsmath}
\usepackage[final]{pdfpages}
\usepackage{pdfpages}

\usepackage[bookmarks=true]{hyperref}
\usepackage{color}

  \hypersetup{
   colorlinks=true,
   pdfborder={0 0 0},
   citecolor=blue,
   linkcolor=blue,
   urlcolor=blue
 }

\begin{document}

\title{$Ab~initio$ calculations for Gamow-Teller strengths in $sd$ shell}


\author{Archana Saxena$^{1}$\footnote{archanasaxena777@gmail.com}, Praveen C. Srivastava$^{1}$\footnote{pcsrifph@iitr.ac.in} and Toshio Suzuki$^{2,3}$\footnote{suzuki@chs.nihon-u.ac.jp}}

\address{$^{1}$Department of Physics, Indian Institute of Technology Roorkee,
Roorkee 247 667, India}

\address{$^{2}$Department of Physics, College of Humanities and Science, Nihon Univerity, Sakurajosui 3, Setagaya-ku, Tokyo 156-8550, Japan}

\address{$^{3}$National Astronomical Observatory of Japan, Osawa 2, Mitaka, Tokyo 181-8588, Japan}

\date{\hfill \today}

\begin{abstract}

In the present work we perform a systematic shell model study of
Gamow-Teller transition strength distributions in
$sd$ shell nuclei using $ab~initio$ effective interactions. The $ab~initio$ effective interactions are based on in-medium similarity renormalization group (IM-SRG)
 and coupled-cluster effective interaction (CCEI) approaches. 
 The aim of the present work is to test the predictive power of $ab~initio$ effective interactions by using the available experimental data of Gamow-Teller strength distributions
in $sd$ shell nuclei. We perform calculations for $^{20}$Ne  $\rightarrow$ $^{20}$F, $^{23}$Na
$\rightarrow$ $^{23}$Mg, $^{23}$Na $\rightarrow$ $^{23}$Ne,
$^{24}$Mg $\rightarrow$ $^{24}$Na, $^{24}$Mg $\rightarrow$ $^{24}$Al, $^{25}$Mg $\rightarrow$ $^{25}$Al, $^{26}$Mg $\rightarrow$ $^{26}$Na, $^{26}$Mg $\rightarrow$ $^{26}$Al,
$^{26}$Si $\rightarrow$ $^{26}$Al, $^{27}$Al $\rightarrow$ $^{27}$Si, $^{28}$Si $\rightarrow$ $^{28}$P, $^{31}$P $\rightarrow$ $^{31}$Si,
and $^{32}$S $\rightarrow$ $^{32}$P
transitions. For comparison we also show the results obtained by using the phenomenological USDB Hamiltonian.
 The  phenomenological USDB results of the Gamow-Teller (GT$_+$/GT$_-$) strength distributions 
show reasonable agreements with the experimental data in comparison to the $ab~initio$ interactions.
 We also calculate the electron capture reaction rates for 
$^{23}$Na(e$^-$, $\nu$)$^{23}$Ne and $^{25}$Mg(e$^-$, $\nu$)$^{25}$Na using $ab~initio$ and USDB interactions.

\end{abstract}


\maketitle

\section{Introduction \label{intro}}
  The Gamow-Teller (GT) transitions are one of the important tools to explore the structure of atomic nuclei \cite{
gt_spoves,gt_cole, Sarriguren,abinitio,pcsepja, pcsjpg}. 
It has many applications such as $\beta$-decay in stellar evolution and electron capture \cite{Langanke2003,Fujita2011,Balasi2015,gt_suzuki};
double electron capture for heating of stars \cite{Martinez2014} and neutrino nucleosynthesis \cite{Woos,Heger,SK,Byelikov}.
There are two types of GT transitions, the GT$_+$ transition where a proton changes into a neutron, and the GT$_-$ transition where a neutron changes into a proton. 
The experimental $B(GT)$ strengths can be obtained from $\beta$-decay studies, but
the excitation energies are limited by the decay $Q$-values in this approach. On the other hand, with the charge exchange (CE) reactions, such as $(p,n)$, $(n,p)$, ($d$,$^{2}$He)
or  ($^{3}$He,$t$), it is possible to access GT transitions at higher energies without the $Q$-value limitation.
The experimental measurements at scattering angles around 0$^{\circ}$ and incident energies above 100 MeV/nucleon provide valuable information on the GT transitions.

There are various experimental probes for the measurement of the GT strengths in $sd$ shell nuclei in $A=20-32$ mass region.
For the GT$_+$ transition, the $(n,p)$ and $(d,^{2}$He) reactions are mainly used to 
obtain the strength distribution.
The $(t,^{3}$He) reaction is also an alternative tool. The GT transition strengths extracted from $\beta$-decay and CE
reactions provide also important tests for $ab ~ initio$ calculations.

Modern $ab~initio$ approaches, like the IM-SRG \cite{hergert}, the coupled cluster theory \cite{dean} and the self-consistent
Green's function method \cite{soma}, have been established and
provide accurate description of nuclear structure properties. 
The $ab~initio$ approaches are more fundamental, although in many cases empirical interactions still are used as benchmarks.
The $ab~initio$ calculations can be used not
only for spherical nuclei, but also to predict the ground and excited state energies and deformations for doubly
open-shell nuclei \cite{stroberg}. The yrast  deformed states in $^{20}$Ne and $^{24}$Mg  have been recently reported in Ref. \cite{stroberg}.

The aim of the present study is to calculate the GT strengths for $sd$ shell nuclei using $ab~initio$ approaches.
In Table \ref{tab:summary}, we give a list of $sd$ shell nuclei and corresponding types of reactions, for which we perform the shell model calculations. Thus, the present work will add more information to earlier $B(GT)$ results for $sd$ shell nuclei obtained by using phenomenological interactions.
Previously, the $B(GT)$ strengths for $A= 10-13$ ($p$ shell) nuclei, using the no-core shell 
model with two-nucleon ($NN$) and three-nucleon ($3N$) interactions
derived from chiral effective field theory, were reported in Ref. \cite{PRL990425012007}. 
Results for the GT strength in the $^{7}$Be($3/2_{g.s}^-$) $\rightarrow$ $^{7}$Li($3/2_{g.s}^-$,$1/2_{1}^-$) transition, obtained by  
using the no-core shell model with the Argonne V8$^\prime$ $NN$ potential and 
the Tucson-Melbourne TM$^\prime$(99) $3N$ potential, were reported in  Ref. \cite{PRC68034305}.

In the next section, we discuss details on the $ab~initio$ effective interactions, along with the phenomenological USDB interaction and
the GT operator. We also discuss the Ikeda sum rule to check the predictive power of $ab~initio$ interactions.
In section III, comparison between the experimental and calculated GT strengths for 13  sets of transitions are reported. 
In section IV, we discuss the electron capture rates.
Finally,  a summary of the present work is given in Section V.

\begin{table*}
\begin{center}
\caption{\label{tab:summary} List of the GT transitions studied in this work.  The data types available and the types of
theoretical calculations used are given. In the last
column the references for the data sets, which are used for comparison with the theoretical calculations, are given.}
\begin{tabular}{ccccccccccccc}
\hline
\hline
\\
 S.No. &      Initial   & Final      & $\beta-$decay & ($n$,$p$) & ($d$,$^{2}$He) & ($t$,$^{3}$He) & ($^{3}$He,$t$) &($p$,$n$) &  CCEI &  IM-SRG &  USDB &  Ref.  \\
 
          \hline
          \\

 1. &   $^{20}$Ne($0^{+}$) & $^{20}$F($1^{+}$) &      &$\surd$   &    &       &       &      &   $\surd$ &  $\surd$ & $\surd$ & \cite{20Ne_20F}\\
    
 2. &   $^{23}$Na($\frac{3}{2}^{+}$) & $^{23}$Mg($\frac{1}{2}^{+}$,$\frac{3}{2}^{+}$,$\frac{5}{2}^{+}$) &       &   &      &       & $\surd$     &  &$\surd$ & $\surd$     & $\surd$  & \cite{23Na_23Mg}  \\
    
 3. &    $^{23}$Na($\frac{3}{2}^{+}$) & $^{23}$Ne($\frac{1}{2}^{+}$,$\frac{3}{2}^{+}$,$\frac{5}{2}^{+}$) &      &  $\surd$    &     &      &     &  &    $\surd$ & $\surd$     & $\surd$ & \cite{23Na_23Ne}\\

 4. &    $^{24}$Mg($0^{+}$) & $^{24}$Na($1^{+}$) &       &      &  $\surd$     & $\surd$      &       &     &$\surd$ & $\surd$     & $\surd$  &\cite{24Mg_24Na,24Mg_24Na_2} \\

 5. &   $^{24}$Mg($0^{+}$) & $^{24}$Al($1^{+}$) &      &     &      &       & $\surd$      &  $\surd$   &  $\surd$ & $\surd$     & $\surd$   & \cite{24Mg_24Al,24Mg_24Al_2} \\

6. &    $^{25}$Mg($\frac{5}{2}^{+}$) & $^{25}$Al($\frac{3}{2}^{+}$,$\frac{5}{2}^{+}$,$\frac{7}{2}^{+}$) &      &      &       &       & $\surd$   &    &$\surd$ & $\surd$     & $\surd$  & \cite{25Mg_25Al}\\

7. &  $^{26}$Mg($0^{+}$) & $^{26}$Na($1^{+}$) &       &      & $\surd$    & $\surd$    &       &      &$\surd$ & $\surd$     & $\surd$  & \cite{26Mg_26Al_26Na,NPA577}\\

8. &    $^{26}$Mg($0^{+}$) & $^{26}$Al($1^{+}$) &       &      &       &       & $\surd$      &$\surd$    & $\surd$ & $\surd$     & $\surd$ & \cite{26Mg_26Al_26Na,26Mg_26Al} \\

9. &   $^{26}$Si($0^{+}$) & $^{26}$Al($1^{+}$) & $\surd$      &     &       &      &       &     &  $\surd$ & $\surd$     & $\surd$ &  \cite{26Si_26Al}\\
    
10. &     $^{27}$Al($\frac{5}{2}^{+}$) & $^{27}$Si($\frac{3}{2}^{+}$,$\frac{5}{2}^{+}$,$\frac{7}{2}^{+}$) &       &      &       &       &$\surd$      &      &$\surd$ & $\surd$     & $\surd$ & \cite{27Al_27Si} \\
    
11. &    $^{28}$Si($0^{+}$) & $^{28}$P($1^{+}$) &      &      &      &    & $\surd$    & $\surd$     &$\surd$ & $\surd$     & $\surd$ &   \cite{28Si_28P,24Mg_24Al}\\
    
12. &    $^{31}$P($\frac{1}{2}^{+}$) & $^{31}$Si($\frac{1}{2}^{+}$,$\frac{3}{2}^{+}$) &      &  $\surd$   &           &   &  &   &$\surd$ & $\surd$ & $\surd$ & \cite{31P_31Si}\\

13. &    $^{32}$S($0^{+}$) & $^{32}$P($1^{+}$) &     &       & $\surd$   &      &       &      &$\surd$ & $\surd$     & $\surd$ & \cite{32S_32P}\\
    \hline
    \hline
\end{tabular}
\end{center}
\end{table*}

\section{Details on $ab~initio$ calculations }
In order to describe the measured GT strength distribution in $sd$ shell nuclei, we have performed shell model
calculations with two modern $ab ~ initio$ approaches: The in-medium similarity renormalization group (IM-SRG) 
\cite{stroberg} and the coupled cluster effective interaction (CCEI) \cite{jan1,pcs_prc,archana_prc}. 
For comparison, we have also performed calculations with the phenomenological USDB effective interaction \cite{usdb}.
For the diagonalization of matrices we used the shell-model code NuShellX \cite{Nushellx}. 

Stroberg $\it {et ~ al.}$ \cite{stroberg} derived a mass-dependent Hamiltonian for $sd$ shell nuclei by using the
IM-SRG \cite{bonger} based on chiral two- and three-nucleon interactions.
In this method, an initial Hamiltonian $H$, which is normal ordered with respect to a finite-density reference state $|\Phi\rangle$ (e.g., the Hartree-Fock ground state) is given as:  
\begin{multline}
H = E_0 +  \sum_{ij}f_{ij}\{a_i^{\dagger}a_j\}
+ \frac{1}{2!^2}\sum_{ijkl}\Gamma_{ijkl}\{a^{\dagger}_ia^{\dagger}_ja_la_k\}\\
+ \frac{1}{3!^2}\sum_{ijklmn}W_{ijklmn} \{a^{\dagger}_ia^{\dagger}_ja^{\dagger}_ka_na_ma_l\},
\label{H}
\end{multline}

\noindent where, $E_0, f_{ij}, \Gamma_{ijkl}$ and $W_{ijklmn}$ are the normal ordered zero-, one-, two-, and three- body terms, 
respectively \cite{tsukiyama}.
The normal ordered strings of creation and annihilation operators obey 
$\langle\Phi|\{a_i^{\dagger}\ldots a_j\}|\Phi\rangle = 0$.
Now, a continuous unitary transformation is applied to the Hamiltonian of Eq. \ref {H}. This unitary transformation is 
parameterized
by a parameter $s$ which is called flow parameter:

\begin{equation}
  H(s)=U(s)H U^{\dag}(s)\,\equiv H^{d}(s)+H^{od}(s).
\end{equation}

Here $H^{d}(s)$ is the diagonal part and $H^{od}(s)$ is
the off-diagonal part of the Hamiltonian. 
As $s\rightarrow\infty$, the off-diagonal matrix 
elements become zero.

The evolution of Hamiltonian with the flow parameter $s$ is given as:
\vspace{-0.5mm}
\begin{equation}
\frac{dH(s)}{ds}=[\eta(s),H(s)],
\end{equation}
\vspace{-0.5mm}
\noindent where $\eta(s)$ is the anti-Hermitian generator of the unitary transformation given by
\begin{equation}
\eta(s)\equiv\frac{dU(s)}{ds}U^{\dag}(s).
\end{equation}

The $H^{od}(s)$ permits us to decouple the $sd$ valence space from the core and higher shells as $s\rightarrow\infty$. 
The resulting Hamiltonian is used in the shell model calculations.
In the present calculations, we use the effective interactions with $\hbar\Omega$=24 MeV.

\begin{figure}
\centering
\includegraphics[width=9.6cm]{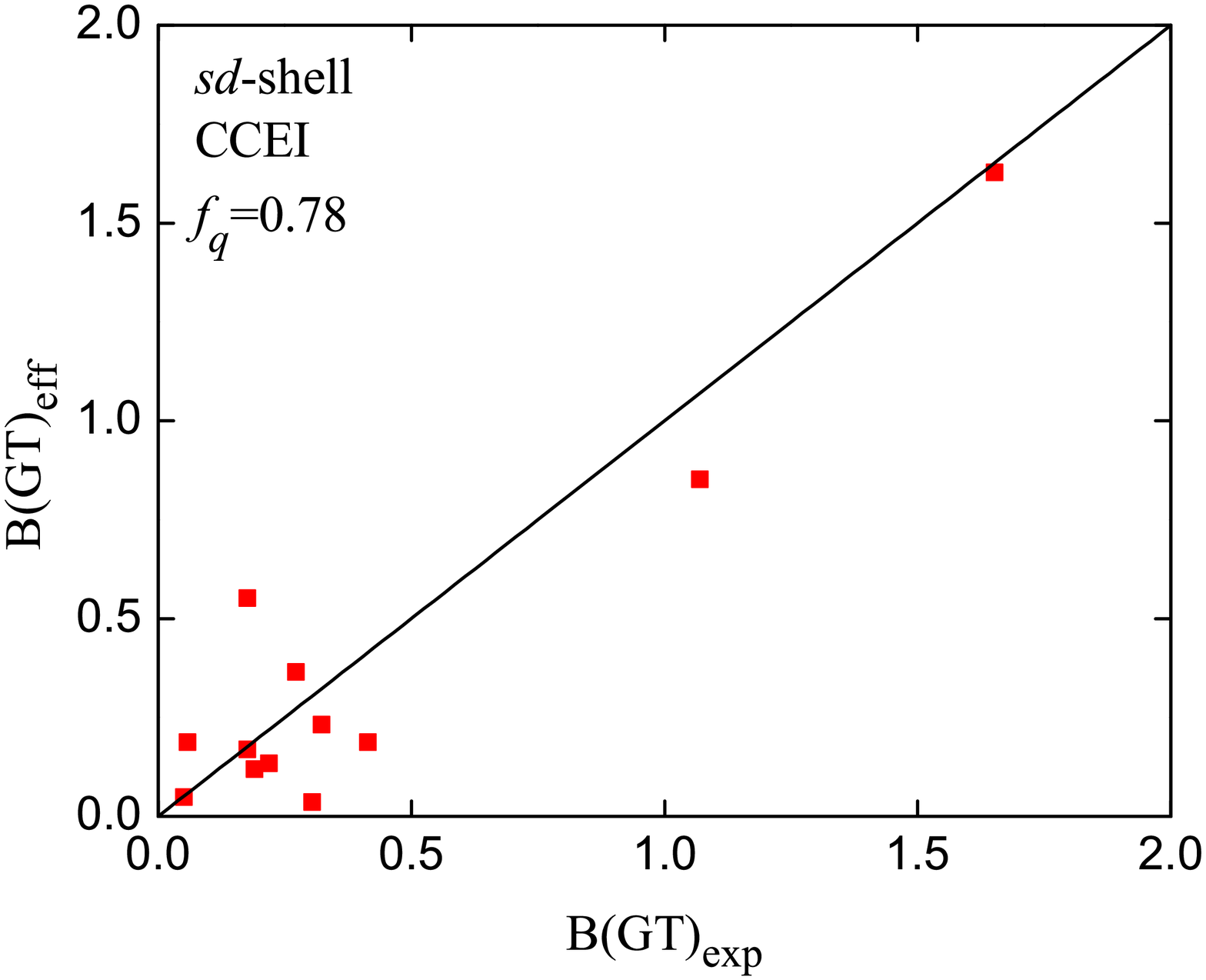} 
\includegraphics[width=9.6cm]{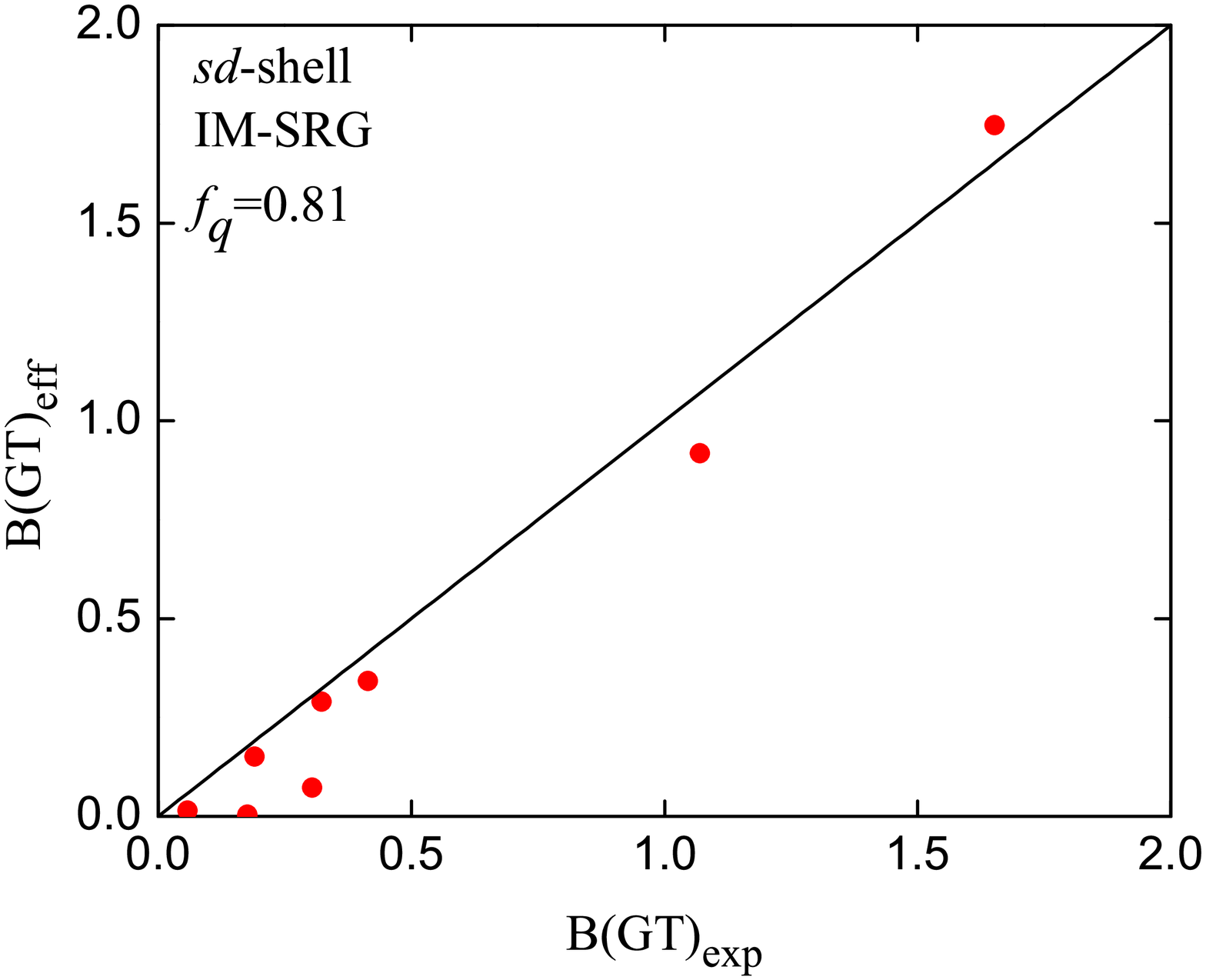} 
\includegraphics[width=9.6cm]{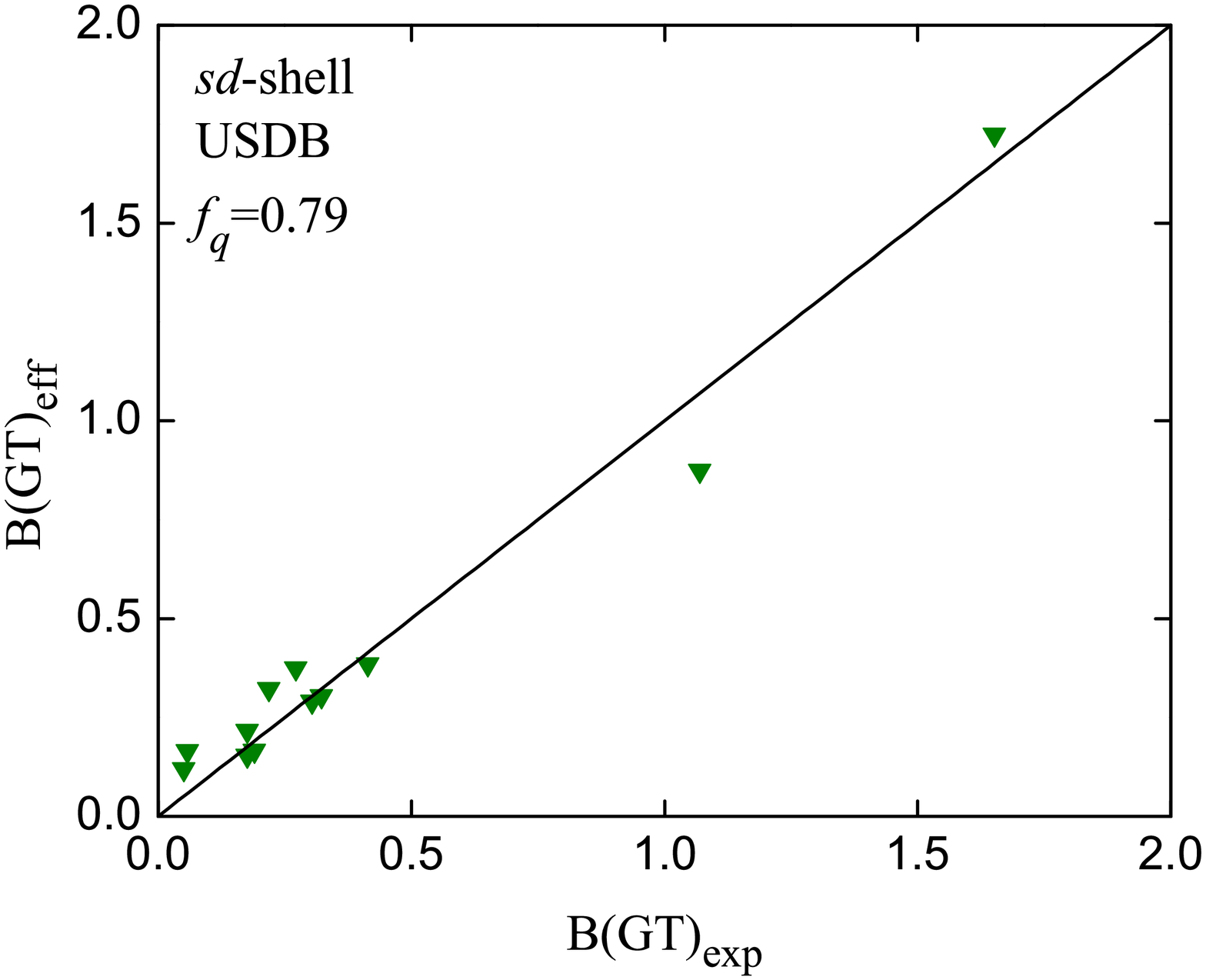} 
\caption{\label{qf}
The calculated value of quenching factor  for  GT transitions in $T=1/2$ $sd$ shell  mirror nuclei  with $A=17-39$ ($A=17-33$) 
using CCEI and USDB (IM-SRG) effective interactions.}
\end{figure}

For the Hamiltonian of the CCEI approach, we have used the following $A$-dependent Hamiltonian
as a starting point: 

\begin{equation}
  \label{intham}
  \hat {H}_{A} = \sum_{i<j}\left({({\bf p}_i-{\bf p}_j)^2\over 2mA} + \hat{V}
    _{NN}^{(i,j)}\right) + \sum_{ i<j<k}\hat{V}_{\rm 3N}^{(i,j,k)}.
\end{equation}

The $NN$ and $3N$ parts are taken from a next-to-next-to-next-to  leading order ($N3LO$) chiral nucleon-nucleon interaction,
and a next-to-next-to leading order ($N2LO$) chiral three-body interaction, respectively.

For both IM-SRG and CCEI, we use $\Lambda_{NN}$= 500 MeV for the chiral $N3LO$ $NN$  interaction \cite{enterm1,enterm2}, and  $\Lambda_{3N}$= 400 MeV for the chiral $N2LO$ $3N$ interaction \cite{piter}.

One can perform a unitary transformation of the Hamiltonian (\ref{intham}) to obtain the
Hamiltonian used for the actual shell model calculations in the CCEI approach: 

\begin{equation}
{H^{A}_{CCEI}}=H_0^{A_c}+H_1^{A_c+1}+H_2^{A_c+2}+ \dots \,.
\end{equation}

Here the first term $H_0^{A_c}$ stands for the core, the second term $H_1^{A_c+1}$ for the 
valence one-body, and $H_2^{A_c+2}$ for the two- body Hamiltonian.
The two-body term is derived from (\ref{intham})
by using Okubo-Lee-Suzuki (OLS) similarity transformation \cite{Okubo1,Okubo2}. 
By applying this unitary transformation,
we get a non-Hermitian effective Hamiltonian.
The similarity transformation is determined from the
metric operator $[S^{\dag}$$S]$= $P_{2}(1+\omega^{\dag}\omega)P_{2}$ (see Ref. \cite{piter2} for further details).
Here, for making the Hamiltonian Hermitian, the metric operator $S^{\dag}$$S$ is used.
The Hermitian shell-model Hamiltonian is then obtained as 
$[S^{\dag}S]^{1/2}\, \hat H^{A}_{CCEI} \, [S^{\dag}S]^{-1/2}$.  

In the present work two-body matrix elements for IM-SRG and CCEI $ab ~ initio$ approaches have been adopted from 
Refs. \cite{stroberg} and \cite{jan1}, respectively.


The Gamow-Teller strength  $B(GT)$ is calculated using the following expression,
\begin{equation}
 {B(GT_{\pm})} = \frac{1}{2J_i + 1} f_q^2 \, |{\langle {f}|| \sum_{k}{\sigma^k\tau_{\pm}^k} ||i \rangle}|^2,
\end{equation}
where  $\tau_+|p\rangle = |n\rangle$, $\tau_-|n\rangle  = |p\rangle$, the index $k$ runs over the single particle orbitals, 
and $|i \rangle$ and $|f \rangle$ describe the state of the parent and daughter nuclei, respectively.
In the present work we have taken the value of quenching factor as $f_q = 0.77$ \cite{brown_review,richter2008}.
 In order to support our above choice, we show the calculated quenching factors for $T=1/2$ $sd$ shell nuclei with 
$A=17-39$ \cite{brown_prc} using the three different interactions in  Fig. \ref{qf}.  
These quenching factors are obtained by a chi-square fit of the theoretical GT transition strengths to the corresponding experimental strengths.
 Note that the IM-SRG interaction and corresponding theoretical transition strengths are available up to $A =34$.  
We can see that the quenching factors obtained are 0.79, 0.78 and 0.81 for
the USDB, CCEI and IM-SRG interactions, respectively.
Although these values show a slight dependence on the interaction, in this work we adopt $f_q=0.77$ for all three interactions, as this is more consistent with the value $f_q = 0.764\pm0.013$ obtained for USDB in ref. \cite{richter2008}, where more data have been used for the fitting.
With this choice $f_q = 0.77$, the r.m.s. deviations from the experimental values are 
 0.088, 0.177 and 0.149 for USDB, CCEI and IM-SRG interactions, respectively.
Compared to the USDB case, the enhancement of the r.m.s. deviations for CCEI and IM-SRG is qualitatively similar to the deviations for
the energy levels \cite{stroberg,jan1}.
In the case of IM-SRG, the calculated $B(GT)$ values are very small and the deviations from the experimental values become large for higher mass nuclei with $A\ge$27.
In the case of CCEI, large deviations are also seen in several nuclei with higher mass, $A$ = 31, 27, 25 and 33 with descending order  of magnitude. This tendency may be attributed to the increasing number of 3-valence nucleon combinations interacting via 
$3N$ forces\cite{stroberg}, which we neglected in our calculation.

In the present work we have also checked the Ikeda sum rule ($B(GT_{-})-B(GT_{+}) = 3 (N-Z) $) for $A$ = 23, 24 and 26. 
Both $ab~intio$ interactions used in the IM-SRG and CCEI methods satisfy this sum rule, as does the phenomenological USDB interaction. 
Thus we are confident that  
enough excited states are taken into account in the calculation of GT strengths for $sd$ shell nuclei in the 
two $ab~initio$ calculations.

\section{Comparison of the experimental and theoretical GT strength distributions}

In this section we compare the theoretical results with the experimental data. 
 
\begin{figure}
\begin{center}
\includegraphics[width=9.5cm]{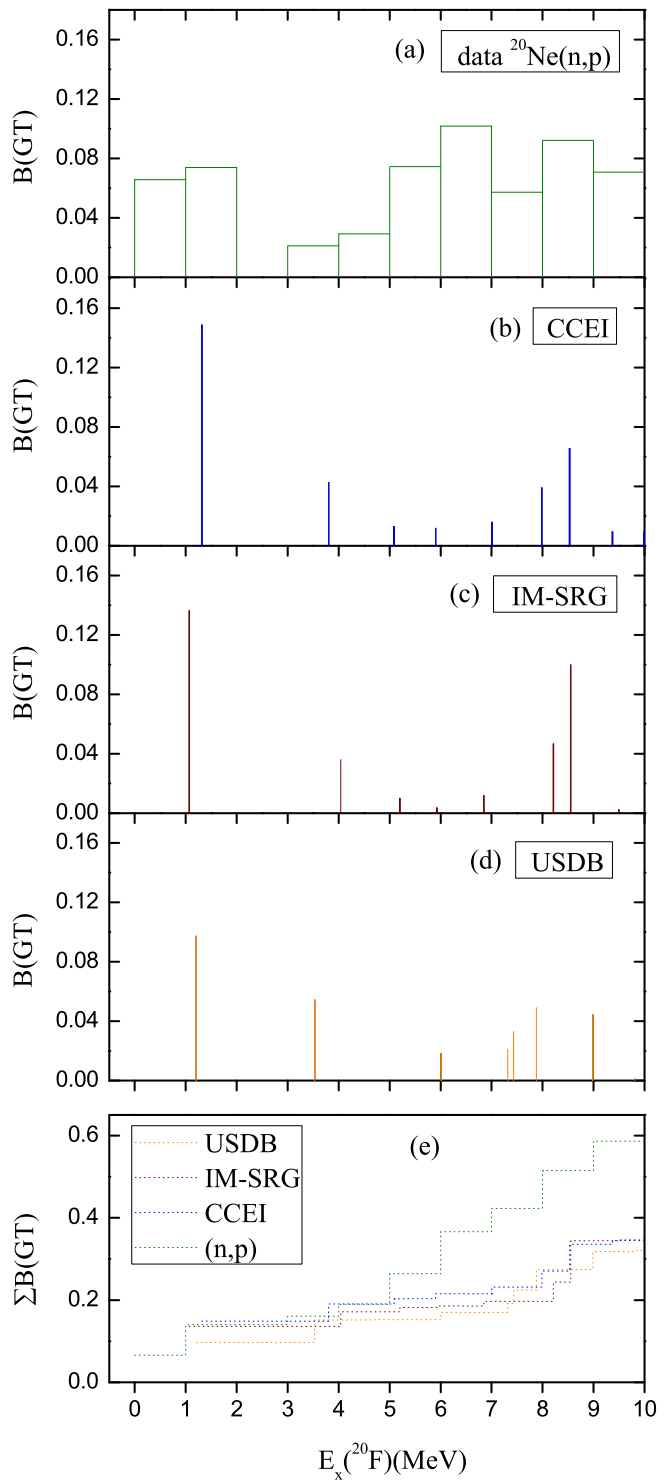} 
\end{center}
\caption{
Comparison of the experimental and theoretical $B(GT)$ distributions for $^{20}$Ne $\rightarrow$ $^{20}$F.}
\label{20Ne_20F}
\end{figure}

\subsection{ $^{20}$Ne $\rightarrow$ $^{20}$F }

In Fig. \ref{20Ne_20F}, we compare the $B(GT)$ strength distribution obtained from the two $ab ~ initio$ effective interactions CCEI (Fig. \ref{20Ne_20F} (b)) and IM-SRG (Fig. \ref{20Ne_20F} (c))  
and the phenomenological USDB (Fig. \ref{20Ne_20F} (d)) interaction 
with the experimental data  
for the transition $^{20}$Ne $\rightarrow$ $^{20}$F. The $B(GT)$ values for these transitions are known from  Ref. \cite{20Ne_20F}, 
where the data is  obtained from the reaction $^{20}$Ne($n$,$p$)$^{20}$F up to the excitation energy $E_{x}$=10 MeV of $^{20}$F. On the  horizontal axis  the excitation energies of
different $1^{+}$ states of 
$^{20}$F are shown. In the experimental data the $B(GT)$ strength is spread over a wide range of excitation energies of $^{20}$F. 
The theoretical calculations for the $B(GT)$ strength 
have already  been done \cite{20Ne_20F} in the framework of the shell model using the universal $sd$ shell (SD) interaction of Wildenthal \cite{USD}.
All three interactions used here give the strongest peak around the excitation energy $\sim$ 1 MeV, and other strong peaks are observed around excitation energies $\sim$ 4 MeV and $\sim$ 8 MeV. The other peaks are small in strength.
In Fig. \ref{20Ne_20F} (c), the strongest peak is observed around the excitation energy $\sim$ 1 MeV, but the strength
is less than that of the strongest peak in the CCEI.
The USDB interaction also shows the strongest peak at $\sim$ 1 MeV, but its strength is smaller than that obtained with both  
$ab ~ initio$ interactions. 
In the experimental data shown in Fig. \ref{20Ne_20F} (a), a wide spread of the $B(GT)$ strength distribution is observed in the
energy range 6-7 MeV and 8-9 MeV.
However, theoretically very small $B(GT)$ strength is obtained in the energy range 6-7 MeV.
All three interactions also show zero strength in the energy range 2-3 MeV.
From Fig. \ref{20Ne_20F}(b), \ref{20Ne_20F}(c) and \ref{20Ne_20F}(d), it is clear that,
as we go towards higher excitation energy, the $B(GT)$ strength decreases and then increases again at the excitation energy 
around 8-9 MeV. 
All three interactions give the ground state (g.s.) $2^{+}$ for $^{20}$F in agreement with the experiment.   
In Fig. \ref{20Ne_20F}(e), the accumulated sums of $B(GT)$ are shown as a function of excitation energy of $^{20}$F.
The CCEI  gives better results compared to the other  
interactions. The summed $B(GT)$ values at higher excitation energies are lower than the experimental data for all 
three interactions used here.
The small calculated $B(GT)$ values in comparison to the experimental values at $E_x >$ 5 MeV can be 
 attributed to the limitation of the configuration space within the $sd$ shell. 
The breaking of $^{16}$O core is important for the fragmentation of the GT strength. 
Shell-model calculations can be performed  in $p$-$sd$ model space to include the $B(GT)$ strength beyond $E_x >$ 10 MeV.
The $^{20}$Ne is a well deformed nucleus, and has admixtures of $g$-shell components \cite{suzuki1977}.
We should also keep in mind that the experimental data have rather large errors, as large as 0.209 for the sum of $B(GT)$ \cite{20Ne_20F}.

\begin{figure}
\begin{center}
\includegraphics[width=9.5cm]{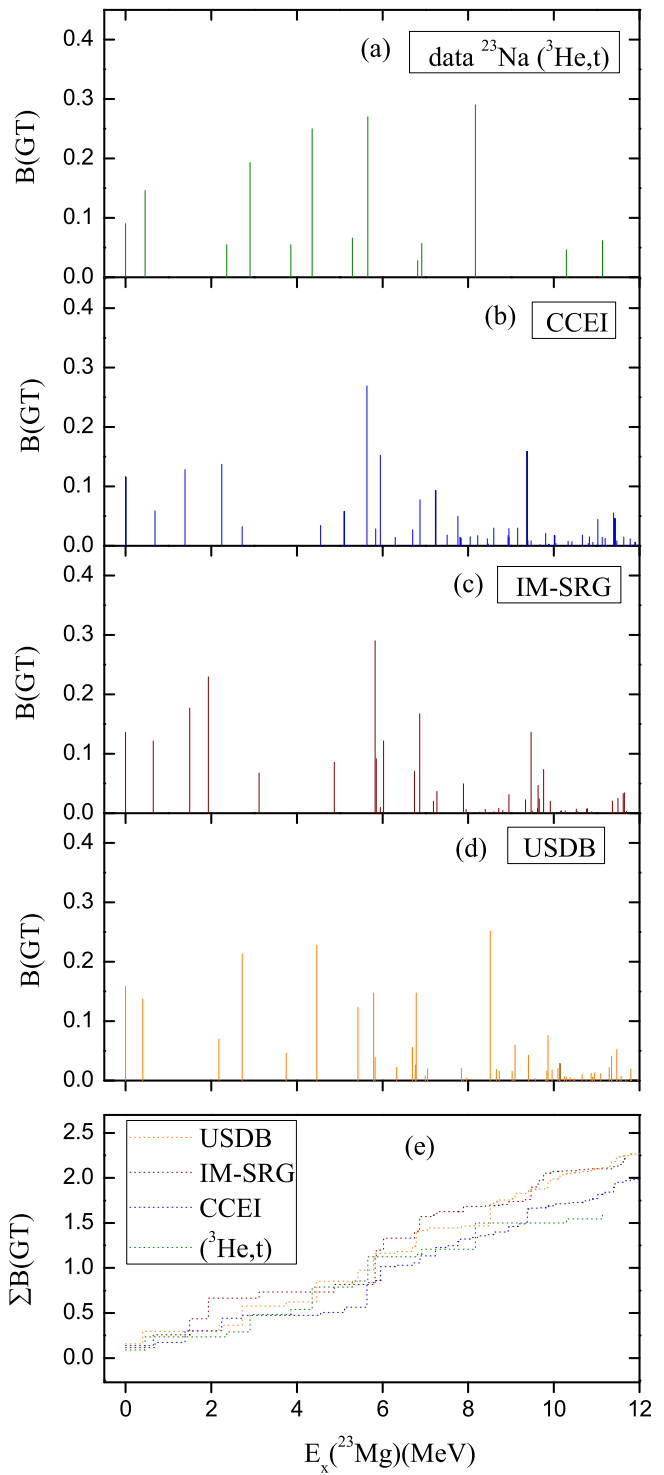} 
\end{center} 
\caption{
Comparison of the experimental and theoretical $B(GT)$ distributions for $^{23}$Na $\rightarrow$ $^{23}$Mg.}
\label{23Na_23Mg}
\end{figure}

\subsection{$^{23}$Na $\rightarrow$ $^{23}$Mg }
Fig. \ref{23Na_23Mg} shows the experimental and theoretical $B(GT)$ strength distribution for the transition 
 $^{23}$Na $\rightarrow$ $^{23}$Mg.
 Here, the experimental data is available from the $^{23}$Na($^{3}$He,$t$)$^{23}$Mg reaction \cite{23Na_23Mg}. 
 In this reaction the $B(GT)$ transitions were measured 
 at the incident energy of 140 MeV per nucleon with the energy resolution of 45 keV. The $^{23}$Na and $^{23}$Mg are deformed nuclei with
 the static quadrupole moments 
 10.1 $\pm$ 0.2 fm$^{2}$ \cite{23Na_23Mg_2} and 11.4 $\pm$ 0.3 fm$^2$ \cite{23Na_23Mg_3}, respectively. These nuclei are
 important from astrophysical point of view.
 In Fig. \ref{23Na_23Mg} (a) the experimental data are shown up to the excitation energy 11.132 MeV of $^{23}$Mg.
 In Ref. \cite{23Na_23Mg}, the strength of the first transition  ($^{23}$Na($\frac{3}{2}^{+}$) $\rightarrow$ $^{23}$Mg($\frac{3}{2}^{+}$)) 
 is 0.340, which is the mixture of $B(GT)$ and Fermi transition strength. 
  We have excluded the Fermi transition strength. Now, the $B(GT)$ transition strength is 0.09.
 The largest $B(GT)$ strength is at the excitation energy 8.168 MeV.
 Fig. \ref{23Na_23Mg} (b) shows the $B(GT)$ strength obtained in the CCEI. 
 The CCEI gives a strong peak at the excitation energy 5.637 MeV, and the magnitude of strength is also
 comparable with the strongest peak observed in the experiment. This peak comes from the transition $^{23}$Na($\frac{3}{2}_1^{+}$) $\rightarrow$
 $^{23}$Mg($\frac{5}{2}_{4}^{+}$). In the energy range 6-12 MeV, we see many $B(GT)$ transitions which
 are not observed in the experiment.
 Fig. \ref{23Na_23Mg} (c) shows results of the IM-SRG interaction. The IM-SRG interaction gives the 
strongest peak of the $B(GT)$ strength at the excitation energy 5.826 MeV,  
 and its strength is comparable with  
 the strength of the strongest peak obtained with the CCEI and observed in the experiment. The density of peaks is smaller
for the IM-SRG interaction compared to the CCEI.
 In Fig. \ref{23Na_23Mg} (d),  we see the strongest
 peak at the excitation energy 8.513 MeV. All three interactions give the g.s. $\frac{3}{2}^{+}$ for $^{23}$Mg,
 in agreement with the experimentally observed g.s. 
 The distribution of accumulated sums of $B(GT)$ for the experimental data and the theoretical calculations is shown in Fig. \ref{23Na_23Mg} (e).
 The CCEI and USDB interactions
 show  similar trends for the summed  $B(GT)$.
 The IM-SRG interaction gives larger values for the summed $B(GT)$ strengths than
 the CCEI or the experiment.

\begin{figure}
\begin{center}
\includegraphics[width=9.5cm]{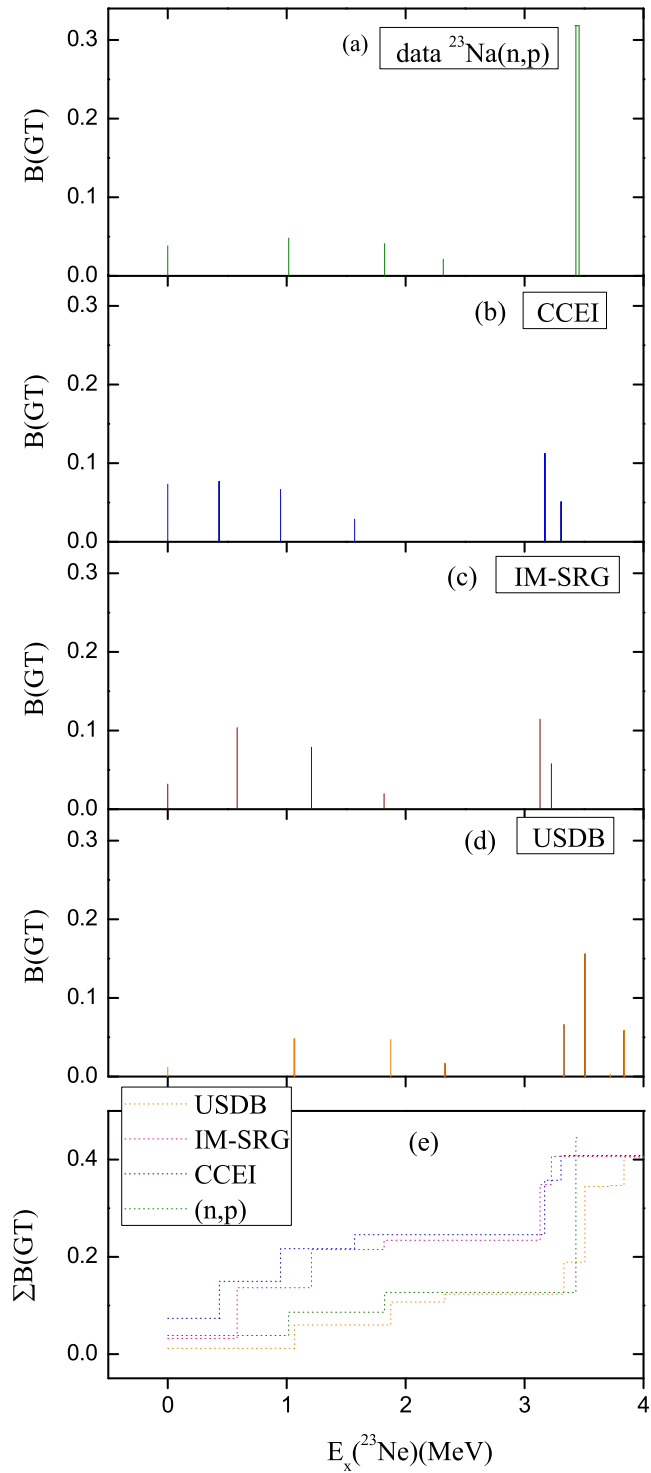} 
\end{center}
\caption{
Comparison of the experimental and theoretical $B(GT)$ distributions for $^{23}$Na $\rightarrow$ $^{23}$Ne .}
\label{23Na_23Ne}
\end{figure}

\subsection{$^{23}$Na $\rightarrow$ $^{23}$Ne }
In Fig. \ref{23Na_23Ne} (a) the experimental $B(GT)$ strength distribution for the transition $^{23}$Na $\rightarrow$ 
$^{23}$Ne \cite {23Na_23Ne}, observed in the charge exchange reaction $^{23}$Na($n,p$)$^{23}$Ne, is shown.
Previously, the shell model results for $B(GT)$ distribution were shown in Ref. \cite{23Na_23Ne}, 
and in the present work we show the calculations obtained by using the recent phenomenological
 USDB interaction in comparison with the $ab~initio$ effective interactions. 
 The experimental $B(GT)$ strength (Fig. \ref{23Na_23Ne} (a)) is dominated in the  excitation energy range 3.432-3.458 MeV. 
The other peaks outside this range have very small $B(GT)$ values. 
 With the CCEI (Fig. \ref{23Na_23Ne} (b)), we get the strongest peak at the excitation energy 3.170 MeV, but
 its strength is approximately three times less than the experimental value.
 We also see some other peaks with $B(GT)$ values below 0.1.
 In the IM-SRG approach (\ref{23Na_23Ne} (c)), we get two peaks with strengths larger than 0.1.
 In Fig. \ref{23Na_23Ne} (d), the  $B(GT)$ distribution is shown for the USDB interaction. 
 We see the strongest peak at the excitation energy 3.508 MeV.
 All strong peaks in the theoretical calculations  correspond to the transition 
 $^{23}$Na($\frac{3}{2}_1^{+}$)$\rightarrow$ $^{23}$Ne($\frac{1}{2}_{2}^{+}$). 
 All three interactions give the g.s. $\frac{5}{2}^{+}$ for $^{23}$Ne, in agreement with the experiment.
  The summed $B(GT)$ values (Fig. \ref{23Na_23Ne} (e)) obtained with USDB show a similar trend as the experimental
values up to the excitation energy 3.5 MeV. 
 The theoretical $B(GT)$ strength is generally lower than the experimental one.

\begin{figure}
\begin{center}
\includegraphics[width=9.5cm]{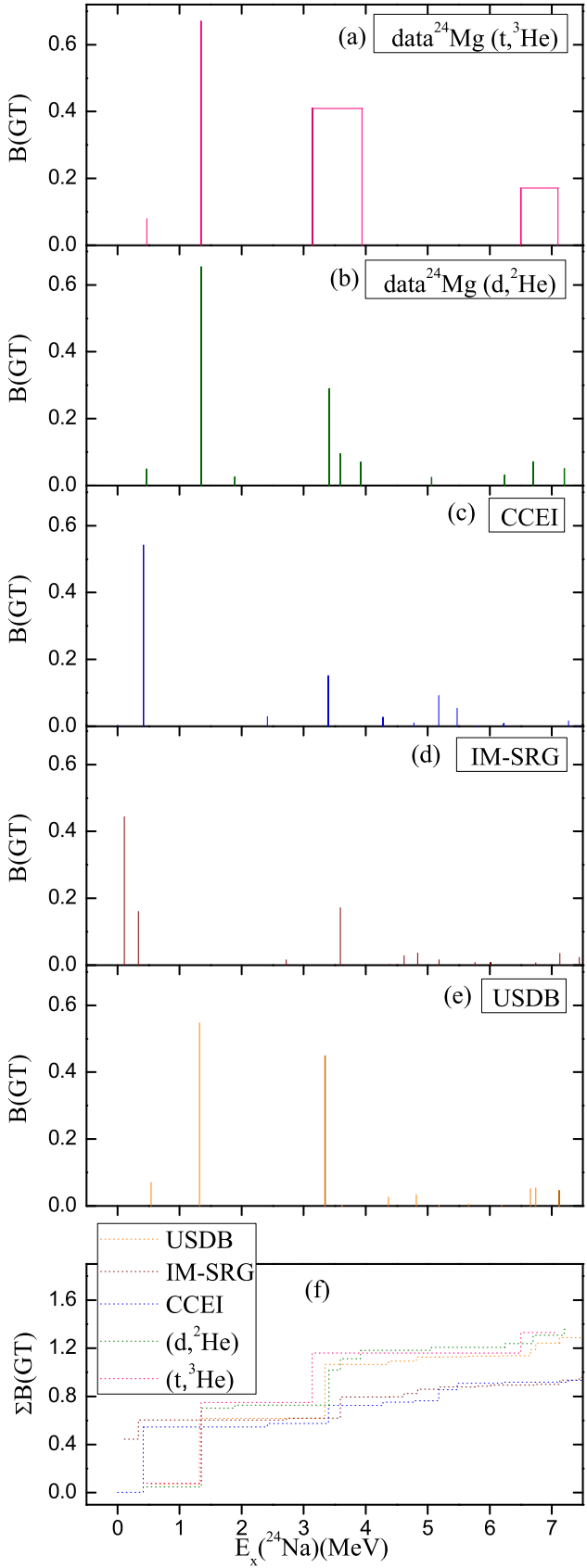} 
\end{center}
\caption{
Comparison of the experimental and theoretical $B(GT)$ distributions for $^{24}$Mg $\rightarrow$ $^{24}$Na.}
\label{24Mg_24Na}
\end{figure}

\subsection{$^{24}$Mg $\rightarrow$ $^{24}$Na }
Figure \ref{24Mg_24Na} shows the $B(GT)$ strength distribution obtained from the shell model calculations,
and the experimental data for the transition $^{24}$Mg $\rightarrow$ $^{24}$Na. 
There are two experimental data for the $B(GT)$ distribution available from the $^{24}$Mg($t$,$^{3}$He)$^{24}$Na
and $^{24}$Mg($d$,$^{2}$He)$^{24}$Na  reactions. 
The $^{24}$Mg($t$,$^{3}$He)$^{24}$Na reaction was performed at the energy of 115 MeV per nucleon, using a
secondary triton beam with the energy resolution of about 200 keV \cite{24Mg_24Na}. 
The $^{24}$Mg($d$,$^{2}$He)$^{24}$Na reaction was performed at the energy 170 MeV,
and a good resolution of the order of 145 keV was obtained in this reaction \cite {24Mg_24Na_2} .
The shell model calculations with the phenomenological interactions USDA and USDB have already  been performed \cite{24Mg_24Na}.
Fig. \ref{24Mg_24Na} (a) shows the data from $^{24}$Mg($t$,$^{3}$He)$^{24}$Na reaction.
In this case, we see the strongest peak at the excitation energy 1.346 MeV of $^{24}$Na. We also find that the 
distribution of the $B(GT)$ strength is in the energy windows from 3.14-3.94 MeV and 6.5-7.1 MeV 
for this reaction.
Fig. \ref{24Mg_24Na} (b) shows the experimental information for the $B(GT)$ distribution from $^{24}$Mg($d$,$^{2}$He)$^{24}$Na
reaction. This reaction gives the 
strongest  peak at the excitation energy 1.35 MeV with nearly the same $B(GT)$ value as obtained from the 
$^{24}$Mg($t$,$^{3}$He)$^{24}$Na reaction.
Some other peaks are also observed with less strength.
Fig. \ref{24Mg_24Na} (c) shows the $B(GT)$ distribution obtained with the CCEI. 
The CCEI predicts $1^{+}$ as the g.s. of $^{24}$Na, while the g.s. from the experiment is $4^{+}$.
In Fig. \ref{24Mg_24Na} (c) we see the strongest peak at the excitation energy 0.417 MeV, which comes from the transition 
$^{24}$Mg($0^{+}$) $\rightarrow$ $^{24}$Na($1_2^{+}$). 
The $B(GT)$ distribution obtained with the IM-SRG interaction is shown in Fig. \ref{24Mg_24Na} (d).
 The IM-SRG interaction gives the g.s. of $^{24}$Na as $2^{+}$.
With this interaction we get the strongest peak at the excitation energy 0.110 MeV, which comes from the 
transition $^{24}$Mg($0^{+}$) $\rightarrow$ $^{24}$Na($1_1^{+}$). Fig. \ref{24Mg_24Na} (e) 
shows the $B(GT)$ distribution obtained with the phenomenological USDB interaction. Here we see  the strongest peak at the excitation
energy 1.323 MeV and the next strongest 
peak at 3.345 MeV, which come from the transitions $^{24}$Mg($0^{+}$) $\rightarrow$ $^{24}$Na($1_2^{+}$)
and $^{24}$Mg($0^{+}$) $\rightarrow$ $^{24}$Na($1_3^{+}$),
respectively.
Fig. \ref{24Mg_24Na} (f) shows the trend of the accumulated sums of the $B(GT)$ distribution obtained from the experimental data
and the theoretical calculations. 
The USDB interaction gives a similar trend as the
experimental data for both reactions.

\begin{figure}
\begin{center}
\includegraphics[width=9.5cm]{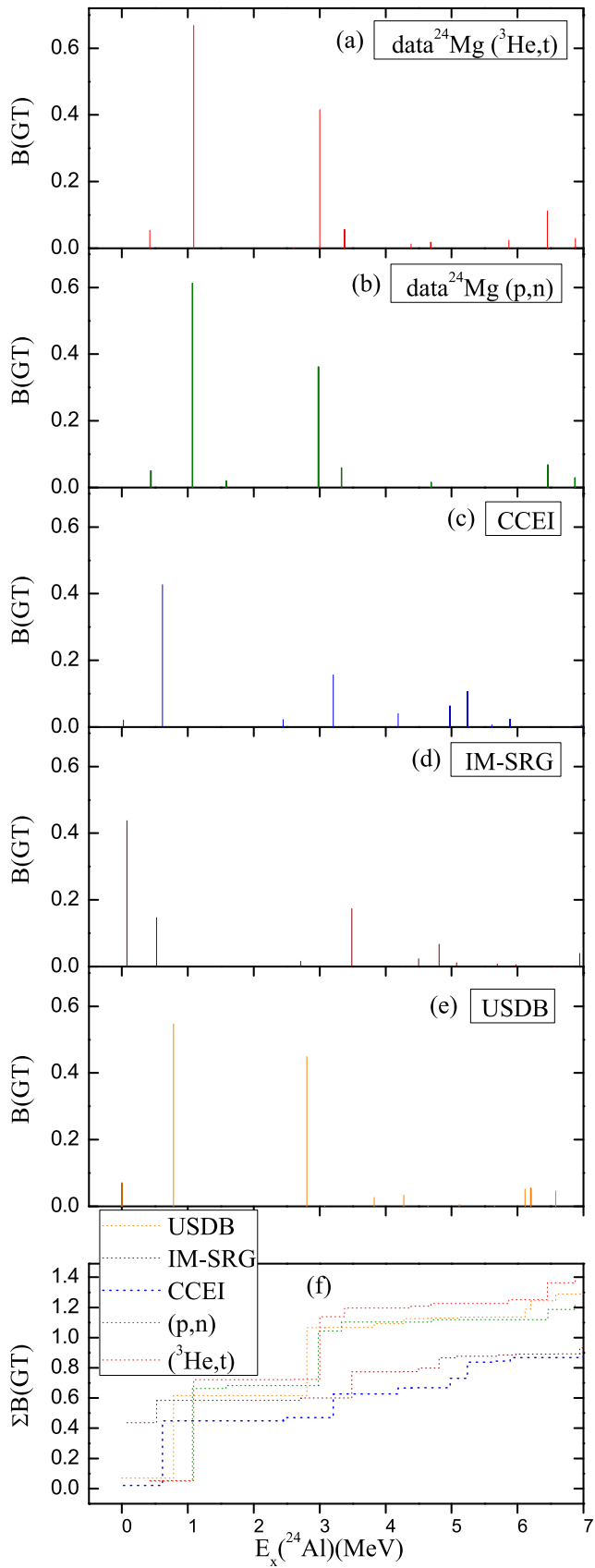} 
\end{center}
\caption{
Comparison of the experimental and theoretical $B(GT)$ distributions for $^{24}$Mg $\rightarrow$ $^{24}$Al.}
\label{24Mg_24Al}
\end{figure}

\subsection{$^{24}$Mg $\rightarrow$ $^{24}$Al}
The experimental information on the $B(GT)$ strength distribution for the transition $^{24}$Mg 
$\rightarrow$ $^{24}$Al is available from the
$^{24}$Mg($^{3}$He,$t$)$^{24}$Al reaction observed at 420 MeV 
\cite{24Mg_24Al_2}, and the $^{24}$Mg($p$,$n$)$^{24}$Al reaction observed at 136 MeV  \cite{24Mg_24Al}.  
The results of the shell model calculation for the $B(GT)$ strength have been previously reported in Ref. \cite{24Mg_24Al_2},
where the phenomenological interactions USDA and USDB were employed.
Fig. \ref {24Mg_24Al} (a) shows the data from the $^{24}$Mg($^{3}$He,$t$)$^{24}$Al reaction.
We see the strongest peak at the excitation energy 1.090 MeV, and  the next strongest peak
at 3.001 MeV. Other peaks are also observed with less strengths. 
Fig. \ref {24Mg_24Al} (b) shows the data from the $^{24}$Mg($p$,$n$)$^{24}$Al reaction.
In this reaction the strongest peak is observed at the excitation energy 1.07 MeV, and the next strongest peak at 2.98 MeV.
In Fig. \ref {24Mg_24Al} (c) the theoretical $B(GT)$ distribution, obtained by using the CCEI, is shown. 
 The CCEI gives the g.s. $2^{+}$ for $^{24}$Al,
whereas the experimental g.s. is $4^{+}$.
The CCEI gives the strongest peak at the excitation energy 0.615 MeV, which comes 
from the transition $^{24}$Mg($0^{+}$)$\rightarrow$ $^{24}$Al ($1_2^{+}$).
The second  strongest peak is observed at the excitation energy 3.205 MeV.
In Fig. \ref {24Mg_24Al} (d) the theoretical $B(GT)$ distribution is shown with the IM-SRG interaction. 
The IM-SRG interaction gives g.s. as  $2^{+}$ for $^{24}$Al.
The IM-SRG interaction gives  the strongest peak at the excitation energy 0.077 MeV which comes 
from the transition $^{24}$Mg($0^{+}$)$\rightarrow$ $^{24}$Al ($1_1^{+}$). The second strongest peak is observed
at the excitation energy 3.486 MeV.
Fig. \ref {24Mg_24Al} (e) shows the  theoretical $B(GT)$ distribution obtained with the USDB interaction. 
The USDB interaction gives the strongest peak at 0.783 MeV which comes 
from the transition $^{24}$Mg($0^{+}$)$\rightarrow$ $^{24}$Al ($1_2^{+}$), and the second strongest peak at
2.805 MeV. The USDB interaction gives the g.s. 
$4^{+}$ for $^{24}$Al, in agreement with the experiment.
The summed $B(GT)$ strength distribution is shown up to the excitation energy 7 MeV. The USDB interaction gives
a reasonable agreement  with the experimental data, while the results 
from $ab ~ initio$ interactions show smaller strength in comparison with the experiment.

\begin{figure}
\begin{center}
\includegraphics[width=9.5cm]{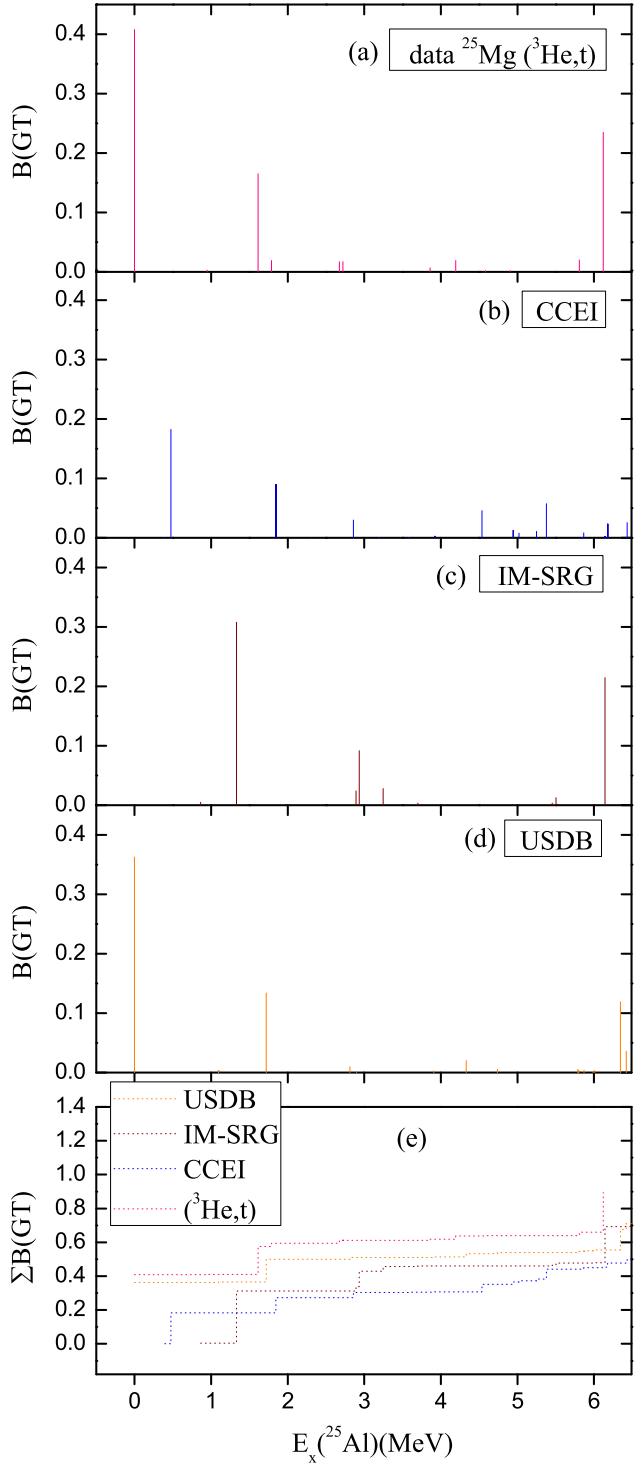} 
\end{center}
\caption{
Comparison of the experimental and theoretical $B(GT)$ distributions for $^{25}$Mg $\rightarrow$ $^{25}$Al.}
\label{25Mg_25Al}
\end{figure}

\subsection{$^{25}$Mg $\rightarrow$ $^{25}$Al }

 The $B(GT)$ strength distribution for the transition  $^{25}$Mg $\rightarrow$ 
$^{25}$Al  is shown in Fig. \ref{25Mg_25Al} (a). 
 This distribution has been measured via the $^{25}$Mg($^{3}$He,$t$)$^{25}$Al reaction at the energy of 140 MeV per nucleon 
\cite{25Mg_25Al}. The $^{25}$Mg and $^{25}$Al nuclei are known to be strongly deformed, and the
 states of these mirror nuclei are well described in terms of the particle rotor model \cite{25Mg_25Al}. 
 In the experiment, the $B(GT)$ strength from the transition $^{25}$Mg($\frac{5}{2}_{1}^{+}$) to  $^{25}$Al($\frac{5}{2}_{1}^{+}$) 
 is dominant, while the other $B(GT)$ strengths are very much suppressed. The explanation of the suppression of $B(GT)$ transitions 
 in $A$ = 25 system is given on the basis of the selection rules of the $K$ quantum number in rotational bands, and also assuming 
the usual selection rule $\triangle J^{\pi}$=$1^{+}$ for  the $B(GT)$ operator. 
  Most of the observed $B(GT)$ transition strength is very small and less reliable, see Ref. \cite{25Mg_25Al}. 
 In particular, it is very weak in the  $\sim$ 2 - 6 MeV energy range. 
  In  Fig. \ref{25Mg_25Al} (b), which shows the theoretical results obtained by using the CCEI approach, we see a 
considerable amount of $B(GT)$ strength in the 2 - 6 MeV energy range. 
 This method gives two dominant peaks at excitation energies 0.474 MeV and 1.847 MeV with smaller $B(GT)$ values 
 than the experiment. The first peak comes from the transition 
 $^{25}$Mg($\frac{5}{2}_{1}^{+}$) to $^{25}$Al($\frac{5}{2}_{1}^{+}$),  and the second
 one  comes from $^{25}$Mg($\frac{5}{2}_{1}^{+}$) to $^{25}$Al($\frac{7}{2}_{1}^{+}$).
In Fig. \ref{25Mg_25Al} (c), which shows the results for the IM-SRG, we see a peak as in the experiment around $\sim$ 6 MeV.
 The strength distribution calculated with the IM-SRG interaction gives the first dominant peak at higher energy in comparison with the experiment.
 Fig. \ref{25Mg_25Al} (d) shows the $B(GT)$ strength distribution obtained with the USDB interaction.
 The first two peaks show a reasonable  agreement with experiment. Above the excitation energy 6 MeV, we see a peak with 
smaller magnitude in comparison with experiment.
 The accumulated sums from the theoretical calculations and experimental data are shown in Fig. \ref{25Mg_25Al} (e).
 The USDB results agree reasonably well with experiment, compared to the 
 $ab~ initio$ interactions.

\begin{figure}
\begin{center}
\includegraphics[width=9.5cm]{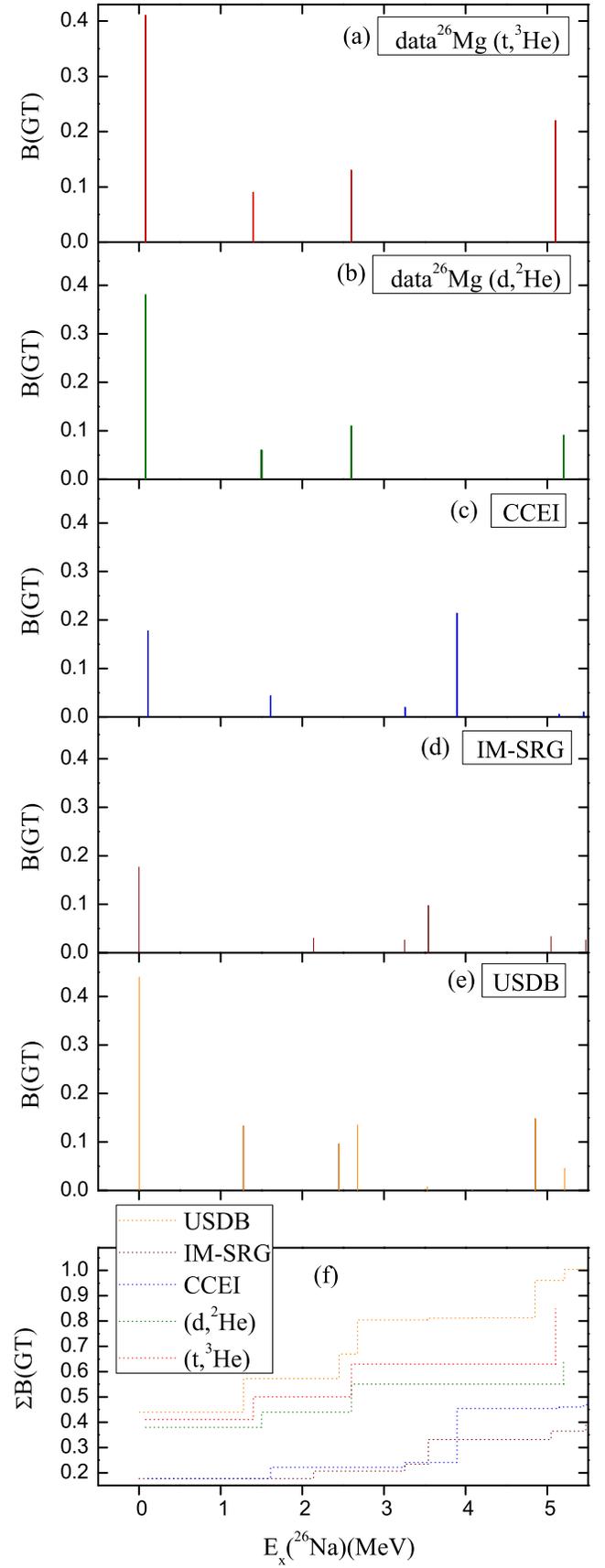} 
\end{center}
\caption{
Comparison of the experimental and theoretical $B(GT)$ distributions for $^{26}$Mg $\rightarrow$ $^{26}$Na.}
\label{26Mg_26Na}
\end{figure}

\subsection{$^{26}$Mg $\rightarrow$ $^{26}$Na }

The $B(GT)$ strength distribution for the transition $^{26}$Mg $ \rightarrow$ $^{26}$Na  is shown in Fig. \ref{26Mg_26Na}.
Fig. \ref{26Mg_26Na} (a) shows  the distribution observed with the  $^{26}$Mg($t$,$^{3}$He)$^{26}$Na reaction at the energy 
of 115 MeV per nucleon \cite {26Mg_26Al_26Na}.
From the experimental data $^{26}$Mg($t$,$^{3}$He)$^{26}$Na, we see the most intense peak at the excitation energy 0.08 MeV of $^{26}$Na. 
Fig. \ref{26Mg_26Na} (b) shows the experimental information from the $^{26}$Mg($d$,$^{2}$He)$^{26}$Na reaction.
This reaction also shows a strong peak at the excitation energy
0.08 MeV. The $B(GT)$ strenghts from the $^{26}$Mg($d$,$^{2}$He)$^{26}$Na data are smaller in magnitude than those from the  $^{26}$Mg($t$,$^{3}$He)$^{26}$Na data.
Fig. \ref{26Mg_26Na} (c) shows the distribution obtained with the CCEI approach. 
This method gives the g.s. of $^{26}$Na as $2^{+}$, whereas the experimental g.s. is $3^{+}$.
With the CCEI, a strong peak is observed at the excitation energy 3.894 MeV.
Fig. \ref{26Mg_26Na} (d) shows the theoretical calculations obtained with the IM-SRG interaction.
From this figure we see the strongest peak at zero excitation energy, but the strength of
this transition is less than half of the strength observed for the strongest peak in the experiment. 
Other calculated strengths are also weak in comparison with both the experimental data.
The IM-SRG interaction gives $1^{+}$ as the g.s. of $^{26}$Na.
Fig. \ref{26Mg_26Na} (e) shows the distribution obtained with the USDB interaction.
It shows a strong peak which is comparable with both experimental data.
The USDB interaction gives $3^{+}$ as the g.s. of $^{26}$Na, in agreement with experiment. 
The accumulated sums are shown in Fig. \ref{26Mg_26Na} (f).
The results obatined from the USDB interaction are much better than the $ab~initio$ interactions.

 \begin{figure}
\begin{center}
\includegraphics[width=9.5cm]{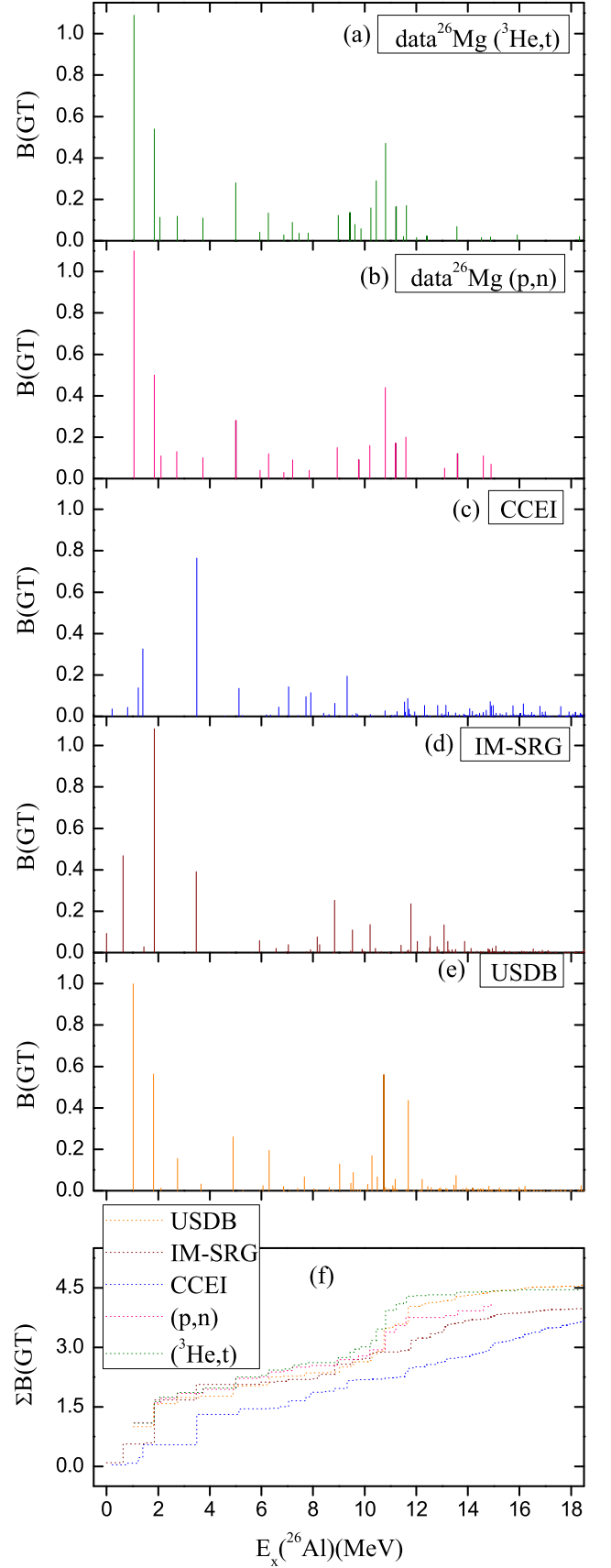} 
\end{center}
\caption{
Comparison of the experimental and theoretical $B(GT)$ distributions for $^{26}$Mg $\rightarrow$ $^{26}$Al.}
\label{26Mg_26Al}
\end{figure}

\subsection{ $^{26}$Mg $\rightarrow$ $^{26}$Al }
Fig. \ref{26Mg_26Al} shows the GT strength distribution for the transition from $^{26}$Mg $ \rightarrow$ $^{26}$Al. 
Information on the $B(GT)$ strength is available from the  $^{26}$Mg($^{3}$He,$t$)$^{26}$Al  \cite {26Mg_26Al_26Na} (Fig. \ref{26Mg_26Al} (a)) 
and $^{26}$Mg($p,n$)$^{26}$Al \cite{26Mg_26Al}  (Fig. \ref{26Mg_26Al} (b)) reactions.
The $^{26}$Mg($^{3}$He,t)$^{26}$Al reaction was observed at 140 MeV/nucleon with energy resolution of 100 keV.
In Fig. \ref{26Mg_26Al} (c) the results obtained by using the CCEI are shown. In this case we see a strong transition at the excitation energy 3.498 MeV of $^{26}$Al, which  comes from the transition $^{26}$Mg ($0^{+}$) $ \rightarrow$ $^{26}$Al ($1_5^{+}$). 
The CCEI gives  the g.s. $3^{+}$  for $^{26}$Al, while the experimental g.s. is $5^{+}$.
Fig. \ref{26Mg_26Al} (d) shows the $B(GT)$ strength distribution using the IM-SRG interaction. 
In this case we see a strong peak at the excitation energy 1.849 MeV, which comes from the transition
$^{26}$Mg ($0^{+}$) $ \rightarrow$ $^{26}$Al ($1_4^{+}$). 
The IM-SRG interaction gives g.s. $1^{+}$ for $^{26}$Al.
In Fig. \ref{26Mg_26Al} (e) the $B(GT)$ strength distribution obtained by using the phenomenological interaction USDB
is shown. This interaction gives a strong peak at the excitation 
energy 1.034 MeV, which comes from the transition $^{26}$Mg ($0^{+}$) $ \rightarrow$ $^{26}$Al ($1_{1}^{+}$).
The USDB interaction gives $5^{+}$ g.s. for $^{26}$Al, in agreement with the experiment. 
Fig. \ref{26Mg_26Al} (f) shows the accumulated sums of 
$B(GT)$ strength for the theoretical calculations and the experimental data. The IM-SRG and USDB interactions show almost the same trend as the experimental data,
 while the CCEI interaction gives lower values.

\begin{figure}
\begin{center}
\includegraphics[width=9.5cm]{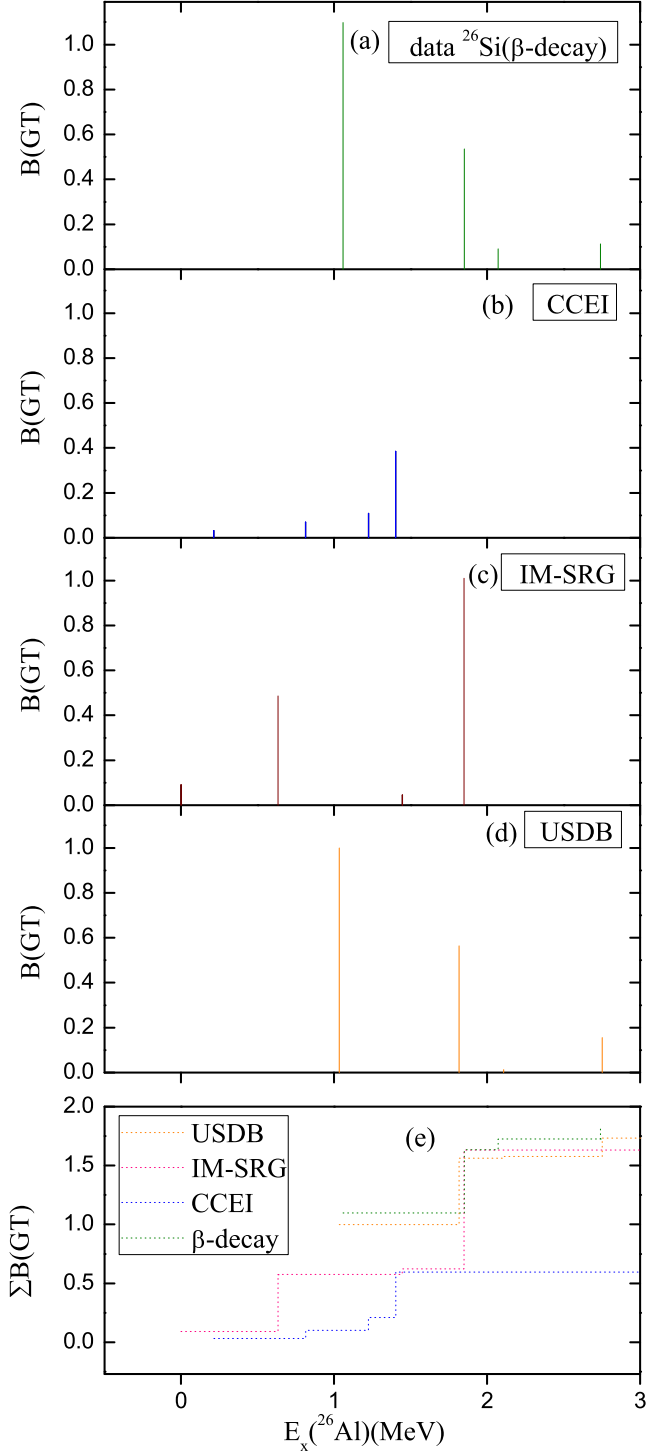} 
\end{center}
\caption{
Comparison of the experimental and theoretical $B(GT)$ distributions for $^{26}$Si $\rightarrow$ $^{26}$Al.}
\label{26Si_26Al}
\end{figure}

\subsection{$^{26}$Si $\rightarrow$ $^{26}$Al }

The experimental and theoretical information on the $B(GT)$ strength distribution for the transition $^{26}$Si $ \rightarrow$ $^{26}$Al is shown in Fig. \ref {26Si_26Al}.
In Fig. \ref {26Si_26Al} (a) the experimental data from $\beta$- decay \cite {26Si_26Al} is shown.
We see a strong peak at the excitation energy 1.0577 MeV of $^{26}$Al. The experimental data are very  sparse. The only four
peaks are observed up to the 
excitation energy 3 MeV of $^{26}$Al. In Fig. \ref {26Si_26Al} (b), the theoretical $B(GT)$ strength distribution 
obtained by using the CCEI is shown. 
It gives a strong peak at the excitation energy 1.403 MeV.  The strength of this peak is smaller
in comparison with the strongest peak from the experimental data. The CCEI gives $3^{+}$ as the g.s.
of $^{26}$Al, whereas the experimental g.s. is $5^{+}$.
Fig. \ref {26Si_26Al} (c) shows the $B(GT)$ strength distribution  for the IM-SRG interaction. We get a strong peak at the
excitation energy 1.849 MeV. 
The strength of this peak is comparable with the experimental data. The IM-SRG interaction gives $1^{+}$  g.s. for $^{26}$Al.
Fig. \ref {26Si_26Al} (d) shows the $B(GT)$ strength distribution from the USDB interaction. The USDB interaction gives a strong peak
at the excitation energy 1.034 MeV.
The results from the USDB interaction are in better agreement with the experimental data, compared to both $ab~initio$ interactions.
The USDB interaction gives $5^{+}$ g.s. for $^{26}$Al, in agreement with the experiment.
Fig. \ref {26Si_26Al} (e) shows the accumulated sums of $B(GT)$ strengths for the experimental data and the theoretical calculations.
The summed $B(GT)$ strength from the USDB interaction matches well with the experimental data, whereas the IM-SRG interaction 
shows the same trend after the 
excitation energy 2 MeV. The CCEI gives lower values in comparison with the experimental data.

\begin{figure}
\begin{center}
\includegraphics[width=9.5cm]{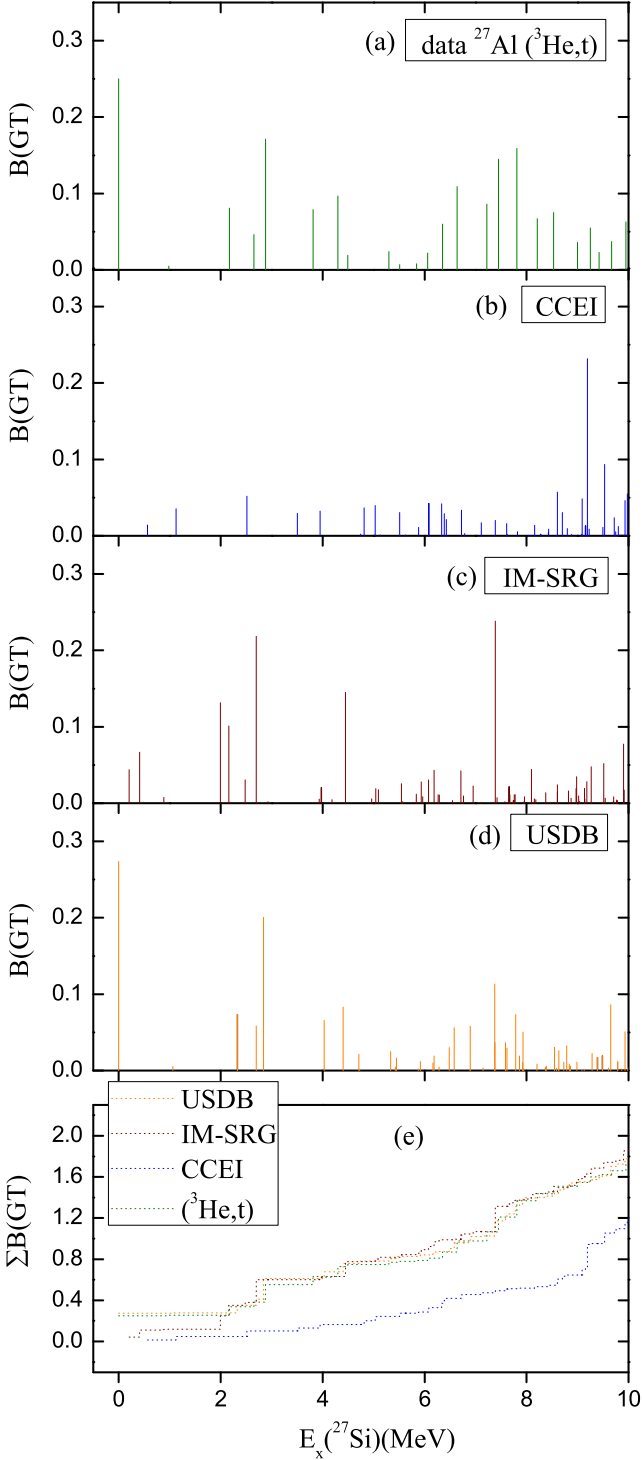} 
\end{center}
\caption{
Comparison of the experimental and theoretical $B(GT)$ distributions for $^{27}$Al $\rightarrow$ $^{27}$Si.}
\label{27Al_27Si}
\end{figure}

 \subsection{\bf$^{27}$Al $\rightarrow$ $^{27}$Si}
 
The nuclei $^{27}$Si and $^{27}$Al are $T$=1/2 mirror nuclei. The information  on the $B(GT)$ strength distribution for the transition 
$^{27}$Al
$\rightarrow$  $^{27}$Si
 is given in Ref. \cite{27Al_27Si}. For these transitions only one experimental data set is available
from the reaction $^{27}$Al($^{3}$He,$t$)$^{27}$Si, which was performed at 150 MeV/nucleon and at scattering angle $0^\circ$. 
The $B(GT)$ strength distribution up to the excitation
energy 9.95 MeV is shown in Fig.  \ref{27Al_27Si} (a). 
In the experimental data, the strength of the transition $^{27}$Al($\frac{5}{2}_{1}^{+}$) $ \rightarrow$ $^{27}$Si($\frac{5}{2}_{1}^{+}$)
is larger than the other $B(GT)$ strengths (At $E_x$($^{27}$Si)= 0.0 MeV,
this strength is obtained by removing the Fermi transition strength).
The experimental $B(GT)$ strength at the excitation energies 0.98, 5.51 and 5.84 MeV 
are not very reliable, and those at 4.49, 5.30 and 6.06 MeV are 
less reliable \cite{27Al_27Si}.  Fig. \ref{27Al_27Si} (b) shows the results obtained with the CCEI. 
At lower energies, the $B(GT)$ strength distribution is very small. We get a strong peak at the excitation energy
9.195 MeV of $^{27}$Si, which  comes from  the transition $^{27}$Al($\frac{5}{2}_1^{+}$) $ \rightarrow$ $^{27}$Si($\frac{7}{2}_{12}^{+}$). 
In the region of the excitation energy range 5-10 MeV, the $B(GT)$ strengths are more dense as compared to below 5 MeV.
For $^{27}$Si, the CCEI gives $\frac{3}{2}^{+}$ as the g.s. of $^{27}$Si, while the experimental 
g.s. is $\frac{5}{2}^{+}$.
The $B(GT)$ strength distribution from the IM-SRG interaction is shown in Fig. \ref{27Al_27Si} (c). The IM-SRG interaction gives 
two strong peaks at energies 2.698 MeV and 7.387 MeV. 
The IM-SRG interaction gives $\frac{3}{2}^{+}$ g.s. for  $^{27}$Si.
Fig. \ref{27Al_27Si} (d) shows the $B(GT)$ distribution from the USDB interaction. In this case we also get two strong
peaks at 0.0 MeV and 2.841 MeV in $^{27}$Si.
The USDB interaction gives $\frac{5}{2}^{+}$ as the g.s. of $^{27}$Si, in agreement with the experiment. 
The comparison of accumulated sums of $B(GT)$ strengths for theoretical
and the experimental values is shown in Fig. \ref{27Al_27Si} (e). 
The USDB and IM-SRG interactions give same trend as
the experimental data, while the CCEI method gives smaller values.

\begin{figure}
\begin{center}
\includegraphics[width=9.5cm]{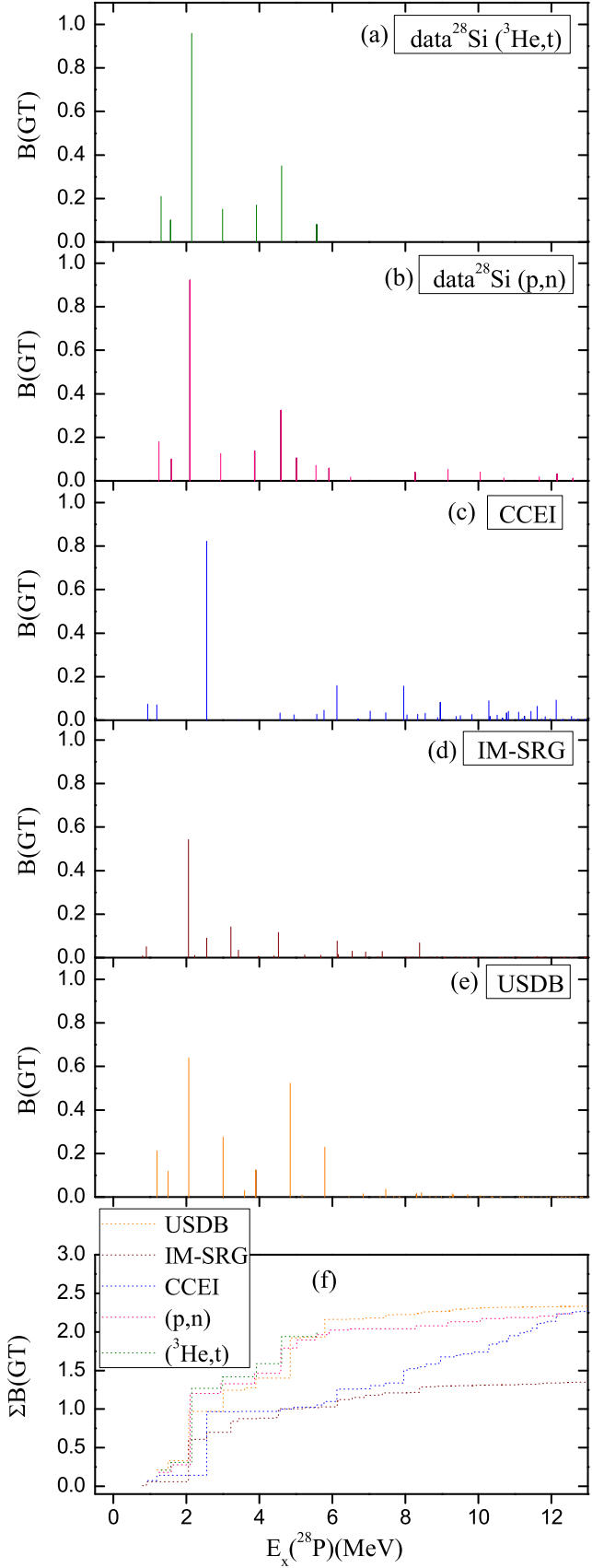} 
\end{center}
\caption{
Comparison of the experimental and theoretical $B(GT)$ distributions for $^{28}$Si $\rightarrow$ $^{28}$P.}
\label{28Si_28P}
\end{figure}


\subsection{$^{28}$Si $\rightarrow$ $^{28}$P }

The experimental information  on the distribution of $B(GT)$ strength is shown in Fig. \ref {28Si_28P}.
There are two experimental data sets available for the transition $^{28}$Si $ \rightarrow$ $^{28}$P.
The charge exchange reaction $^{28}$Si($^{3}$He,$t$)$^{28}$P 
was performed at 150 MeV/nucleon, using the dispersion-matching technique \cite {28Si_28P} to get good energy resolution. 
In Fig.  \ref {28Si_28P}(a) the results for the $^{28}$Si($^{3}$He,$t$)$^{28}$P reaction are shown up to the excitation 
energy 5.57 MeV.  
The shell model study has already been carried out \cite {28Si_28P} by using 
Wildenthal's USD interaction. 
In this figure, a large $B(GT)$  strength is obtained at the excitation energy 2.15 MeV, but this value is
normalized to the ($p$,$n$) data which is taken from 
Ref. \cite {24Mg_24Al}. 
The  $B(GT)$ distribution from the $^{28}$Si($p$,$n$)$^{28}$P reaction performed at energy 136 MeV per nucleon is given
in Fig. \ref {28Si_28P}(b). Here, a large $B(GT)$ strength is obtained at 
the excitation energy 2.10 MeV of $^{28}$P.  Fig. \ref {28Si_28P}(c)
shows the theoretical results by using CCEI. We get a strong transition at 2.562 MeV with $B(GT)$ strength 0.82, which comes from the transition
$^{28}$Si($0^{+}$) $\rightarrow$ $^{28}$P($1_3^{+}$).
We can also see many transitions above the excitation energy 4 MeV of $^{28}$P, but they 
are very small in strength. The experimental g.s. of $^{28}$P is $3^{+}$, while the CCEI predicts $0^{+}$.
Fig. \ref {28Si_28P}(d) shows the $B(GT)$ strength distribution from the IM-SRG interaction. Here, we see a strong peak  
at the excitation energy 2.056 MeV 
of $^{28}$P, which comes from the transition $^{28}$Si($0^{+}$) $\rightarrow$ $^{28}$P($1_3^{+}$). The IM-SRG 
interaction predicts $2^{+}$ as the g.s. of $^{28}$P.
In Fig. \ref {28Si_28P}(e), the $B(GT)$ strength distribution obtained with the USDB interaction is shown. In this case, we
can see two comparable peaks at excitation energies 2.065 and
4.847 MeV of $^{28}$P, which come from the transition $^{28}$Si($0^{+}$) $\rightarrow$ $^{28}$P($1_3^{+}$) and
$^{28}$Si($0^{+}$) $\rightarrow$ $^{28}$P($1_7^{+}$), respectively. The USDB interaction gives $3^{+}$ g.s. for $^{28}$P, in agreement with the experiment. 
In Fig. \ref {28Si_28P}(f), the accumulated sums of $B(GT)$ strengths is shown. The USDB interaction shows 
a similar trend as the experimental data. The IM-SRG interaction gives smaller 
value in comparison with the other interactions and the experimental data.

\begin{figure}
\begin{center}
\includegraphics[width=9.5cm]{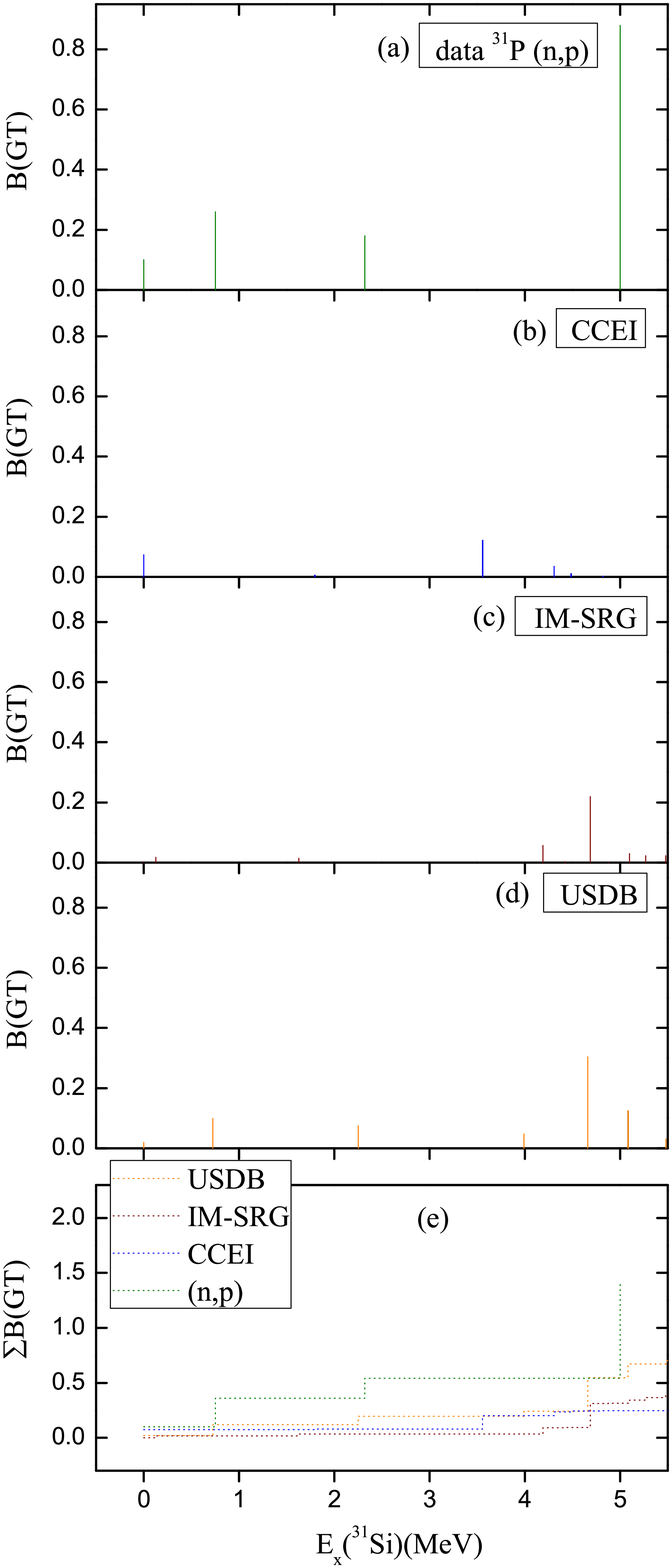} 
\end{center}
\caption{
Comparison of the experimental and theoretical $B(GT)$ distributions for $^{31}$P $\rightarrow$ $^{31}$Si.}
\label{31P_31Si}
\end{figure}

\subsection{$^{31}$P $\rightarrow$ $^{31}$Si }

The $B(GT)$ strength distribution for the transition $^{31}$P $\rightarrow$  $^{31}$Si
is shown in Fig. \ref{31P_31Si}. In Fig. \ref{31P_31Si}(a), the experimental data is shown for 
the reaction $^{31}$P($n,p$)$^{31}$Si \cite{31P_31Si}. 
In the experimental data, we see an intense peak at the excitation energy 5 MeV of $^{31}$Si. 
The charge exchange reaction $^{31}$P($n,p$)$^{31}$Si was performed to find the double
differential cross section with the incident neutron energy of 198 MeV. Using multipole decomposition 
techniques the $B(GT)$ strength distribution was extracted.
The shell model study using the universal $sd$ (USD) interaction has already  been done in Ref. \cite{31P_31Si}.
Fig. \ref{31P_31Si}(b) shows the $B(GT)$ distribution obtained with the CCEI method. 
We see the strongest peak at
the excitation energy 3.557 MeV of $^{31}$Si, which  comes
from the transition $^{31}$P($\frac{1}{2}_{1}^{+}$) $\rightarrow$ $^{31}$Si($\frac{3}{2}_{2}^{+}$). The strength of 
this transition is very small in comparison
to the strength of the strongest peak in the experiment. The CCEI gives $\frac{3}{2}^{+}$ g.s. for $^{31}$Si, which agrees with the experiment.
 In Fig. \ref{31P_31Si}(c), the distribution of
the $B(GT)$ strength is shown for the IM-SRG interaction.
Here, we see  the strongest peak at the excitation 
energy 4.685 MeV of $^{31}$Si, which  comes from the transition $^{31}$P($\frac{1}{2}_{1}^{+}$) $\rightarrow$ $^{31}$Si($\frac{1}{2}_{3}^{+}$),
but also here the strength is very small compared to the strongest peak in the experimental data. 
The IM-SRG also reproduces 
correctly 
the experimental g.s. of $^{31}$Si. Fig. \ref{31P_31Si}(d) shows the $B(GT)$ distribution from the USDB 
interaction. In this case the strongest peak is observed at
the excitation energy 4.661 MeV with strength 0.30, which is larger than the strength of the peaks obtained in 
the $ab~initio$ interactions. This peak  comes from the transition 
$^{31}$P($\frac{1}{2}_{1}^{+}$) $\rightarrow$ $^{31}$Si($\frac{1}{2}_{2}^{+}$). The USDB interaction also
gives the correct g.s. of $^{31}$Si.
Fig. \ref{31P_31Si}(e) shows the accumulated $B(GT)$ strengths for all three interactions and the experimental data. 
All three interactions give small values in comparison to the experimental data.

\begin{figure}
\begin{center}
\includegraphics[width=9.5cm]{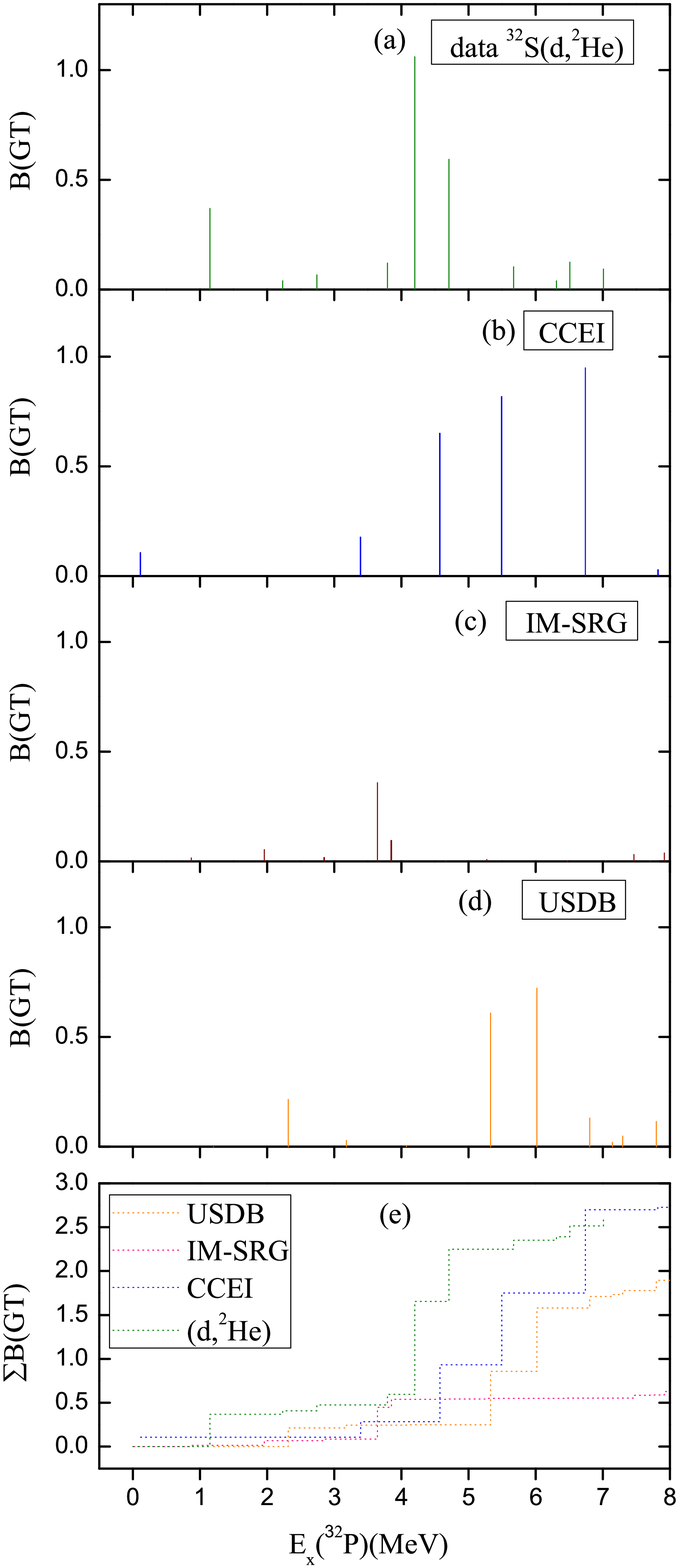} 
\end{center}
\caption{
Comparison of the experimental and theoretical $B(GT)$ distributions for $^{32}$S $\rightarrow$ $^{32}$P.}
\label{32S_32P}
\end{figure}

\begin{table*}
\begin{center}
\caption{\label{tab:centroid} Comparison between the experimental and theoretical centroid energy of GT distributions for $sd$ shell nuclei.}
\begin{tabular}{cccccccccccc}
\hline
\hline
\\

 S.No. &      Initial   & Final      & $\beta-$decay & ($n$,$p$) & ($d$,$^{2}$He) & ($t$,$^{3}$He) & ($^{3}$He,$t$) &($p$,$n$) &  CCEI &  IM-SRG &  USDB   \\
          \hline  
 1. &   $^{20}$Ne($0^{+}$) & $^{20}$F($1^{+}$) &      &5.39   &    &       &       &      &   4.66 &  4.92 & 5.08 \\
    
 2. &   $^{23}$Na($\frac{3}{2}^{+}$) & $^{23}$Mg($\frac{1}{2}^{+}$,$\frac{3}{2}^{+}$,$\frac{5}{2}^{+}$) &       &   &      &       & 5.00   &  &5.86 & 5.19     & 5.59 \\
    
 3. &    $^{23}$Na($\frac{3}{2}^{+}$) & $^{23}$Ne($\frac{1}{2}^{+}$,$\frac{3}{2}^{+}$,$\frac{5}{2}^{+}$) &      &  3.01    &     &      &     &  &    1.63 & 1.81    & 2.10\\

 4. &    $^{24}$Mg($0^{+}$) & $^{24}$Na($1^{+}$) &       &      &  2.73     & 3.22      &       &     &1.96 & 1.67    & 2.82\\

 5. &   $^{24}$Mg($0^{+}$) & $^{24}$Al($1^{+}$) &      &     &      &       & 2.44     &  2.21  &  2.35& 1.50     & 2.28   \\

6. &    $^{25}$Mg($\frac{5}{2}^{+}$) & $^{25}$Al($\frac{3}{2}^{+}$,$\frac{5}{2}^{+}$,$\frac{7}{2}^{+}$) &      &      &       &       & 1.93   &    &2.33 & 2.00     & 0.88 \\

7. &  $^{26}$Mg($0^{+}$) & $^{26}$Na($1^{+}$) &       &      & 1.74   & 1.90    &       &      &2.20 & 1.82    & 1.58  \\

8. &    $^{26}$Mg($0^{+}$) & $^{26}$Al($1^{+}$) &       &      &       &       & 6.08     &5.95   & 8.47 & 6.40    & 6.87\\

9. &   $^{26}$Si($0^{+}$) & $^{26}$Al($1^{+}$) & 1.44     &     &       &      &       &     &  1.24 & 1.37    &1.32 \\
    
10. &     $^{27}$Al($\frac{5}{2}^{+}$) & $^{27}$Si($\frac{3}{2}^{+}$,$\frac{5}{2}^{+}$,$\frac{7}{2}^{+}$) &       &      &       &       &5.35     &      &7.29 & 5.60     & 5.37\\
    
11. &    $^{28}$Si($0^{+}$) & $^{28}$P($1^{+}$) &      &      &      &    & 2.80    & 3.64    &6.08 & 3.82    &3.66\\
    
12. &    $^{31}$P($\frac{1}{2}^{+}$) & $^{31}$Si($\frac{1}{2}^{+}$,$\frac{3}{2}^{+}$) &      &  3.53   &           &   &  &   &2.61 &4.19 & 3.39\\

13. &   $^{32}$S($0^{+}$) & $^{32}$P($1^{+}$) &     &       & 4.10  &      &       &      &5.36 & 3.45    & 5.31\\

  \hline
    \hline
\end{tabular}
\end{center}
\end{table*}


\subsection{\bf$^{32}$S $\rightarrow$ $^{32}$P }
Fig. \ref{32S_32P} presents the experimental and theoretical information on the distribution of the $B(GT)$ 
strength for the transition $^{32}$S $\rightarrow$  $^{32}$P. For the experimental data, the charge exchange reaction $^{32}$S($d$,$^{2}$He)$^{32}$P was
performed at forward angles and at an incident energy of 
$E_{d}$= 170 MeV with a resolution of 150 keV  \cite{32S_32P}.  
Fig. \ref{32S_32P} (a) shows the experimental data for $B(GT)$ strength distribution. Here, we see an intense
peak at the excitation energy 4.2 MeV of $^{32}$P. 
The experimental g.s. of $^{32}$P is $1^{+}$.
Fig. \ref{32S_32P} (b) presents the results obtained by using the CCEI. Here, we see 
three strong peaks at excitation energies 4.573 MeV, 5.494 MeV and 6.741 MeV
in $^{32}$P, which  come from the transitions $^{32}$S($0^{+}$) $\rightarrow$ $^{32}$P($1_{3}^{+}$), 
$^{32}$S($0^{+}$) $\rightarrow$ $^{32}$P($1_{4}^{+}$) and
$^{32}$S($0^{+}$) $\rightarrow$ $^{32}$P($1_{5}^{+}$), respectively. The CCEI predicts the
g.s. of $^{32}$P as $2^{+}$, while the experimental g.s. is $1^{+}$.
Fig. \ref{32S_32P} (c) shows the $B(GT)$ distribution obtained by using the IM-SRG interaction. Here, we notice a 
peak at excitation energy 3.642 MeV of $^{32}$P,
which  comes from the transition $^{32}$S($0^{+}$) $\rightarrow$ $^{32}$P($1_{5}^{+}$), weak in comparison
to the the strongest peak of the experimental data 
and the CCEI results. The IM-SRG interaction give $0^{+}$ as the g.s. of $^{32}$P. 
Fig. \ref{32S_32P} (d) shows the $B(GT)$ distribution obtained by using the phenomenological USDB interaction. 
This interaction gives two peaks with comparable strengths, but weak compared to the 
strongest peak from the experimental data. The strong peaks with the USDB 
interaction are at excitation energies 5.33 MeV and 6.019 MeV of $^{32}$P, which come from the transitions
$^{32}$S($0^{+}$) $\rightarrow$ $^{32}$P($1_{6}^{+}$) and
$^{32}$S($0^{+}$) $\rightarrow$ $^{32}$P($1_{7}^{+}$), respectively.
The USDB interaction gives $3^{+}$ as the g.s. of $^{32}$P.
Fig. \ref{32S_32P} (e) shows the accumulated sums of $B(GT)$ strength. Among the three interactions, the
CCEI gives better results for the accumulated sums.

\subsection{Centroid energies}

In Sects. III.A--III.M, we discussed GT distributions for $sd$ shell nuclei obtained with the two $ab~initio$ 
interactions as well as the USDB.
Among the three interactions, the USDB in general gives the best account of the experimental data.
The $ab~initio$ interactions give rather reasonable account of the experimental data, though there are deviations in many cases. 
In Table \ref{tab:centroid}, we show a comparison between the experimental and theoretical centroid energies of the GT 
distributions for $sd$ shell nuclei.

$Ab~initio$ interactions give larger GT strength than USDB and the experimental data in the lower excitation energy region, for example, 
in the $^{23}$Na $\rightarrow$ $^{23}$Ne and $^{24}$Mg $\rightarrow$ $^{24}$Na ($<$ 1 MeV) transitions.
On the other hand, less GT strength is seen in lower excitation energy region, for example, in the $^{25}$Mg $\rightarrow$ $^{25}$Al and
$^{26}$Mg $\rightarrow$ $^{26}$Na transitions.  
This is also true for the $^{26}$Mg $\rightarrow$ $^{26}$Al and $^{27}$Al $\rightarrow$ $^{27}$Si transitions in case of CCEI.
These differences are reflected in the centroid energies of the GT distributions.
They are smaller (larger) when more (less) strength is found in the lower excitation energy region as shown in Table \ref{tab:centroid}.
In case of the $ab~initio$ interactions, single-particle energy gap between $d_{3/2}$ and $d_{5/2}$ orbits is large compared to USDB, in particular for CCEI.
This could explain the general feature that $ab~initio$ interactions show much lower strength at low energies in higher mass nuclei.  
Especially small GT strengths for CCEI in the transitions shown in Figs. 9-11 can be attributed to insufficient contributions from $d_{3/2}$ orbit due to the largest gap among the interactions. 
As discussed in Sect. II, deviations of the GT strength, calculated by using the $ab~initio$ interactions, from the experimental data become generally larger for the higher mass nuclei because of the lack of three-body cluster terms among valence nucleons.

In the present calculations, we used the one-body GT operator with a universal quenching factor for both $ab~initio$ and phenomenological USDB interactions.
However, the GT operator should be evolved in the same way as the Hamiltonians for the IM-SRG and CCEI methods. 
This gives rise to induced two-body operators \cite{Stroberg2017}. Although this effect is taken into account here by adopting a phenomenological universal quenching factor for the one-body operator, induced two-body operators can lead to mass dependent quenching factors, which may also depend on the interactions. 
The present calculation, therefore, has limitations because of the truncations of the Hamiltonians up to the two-body terms, and the operator up to the one-body term.

\section{Electron capture rates in $^{23}$Na and $^{25}$Mg}

In this section, we apply the GT transition strengths obtained by the $ab~initio$ effective interactions
in $sd$ shell nuclei to evaluate the electron capture rates
in stellar environments.

Electron capture rates at high densities and high temperatures are evaluated as 
\citep{Full80,Full82b,Full82a,Full85}
\begin{equation*}
\lambda = \frac{\rm ln2}{6146 (s)} \sum_{i} W_i \sum_{f} B(GT; i \rightarrow f) 
\end{equation*}
\begin{equation*}
\times  \int_{\omega_{min}}^{\infty} \omega p (Q_{if}+\omega)^2 F(Z, \omega) S_e(\omega) d\omega,  
\end{equation*}
\begin{equation*}
Q_{if} = (M_{\rm p}c^2 -M_{\rm d}c^2 +E_i -E_f)/m_{e}c^2,
\end{equation*}
\begin{equation}
W_i = (2J_i +1) e^{-E_i/kT} /\sum_{i}(2J_i +1) e^{-E_i/kT},
\end{equation}

\noindent where $\omega$ and $p$ are electron energy and momentum
in units of
$m_{e}c^2$ and $m_{e}c$; $M_{\rm p}$ and $M_{\rm d}$ are the masses of parent and daughter nuclei,
and $E_i$ and $E_f$ are excitation energies of initial and final states.
$F(Z, \omega)$ is the Fermi function, and $S_e(\omega)$ is the Fermi-Dirac
distribution for electrons, where the chemical potential ($\mu_{e}$)
is determined from the density ($\rho Y_e$) by
\begin{equation}
\rho Y_e = \frac{1}{\pi^{2}N_{A}}(\frac{m_{e}c}{\hbar})^3
\int_{0}^{\infty}(S_{e} -S_{p})p^2dp.
\end{equation}
Here $N_A$ is the Avogadro number, and $S_{p}$ is the Fermi-Dirac distribution
for positrons with the chemical potential $\mu_{p}$ = $-\mu_{e}$.
Its value can become as large as 2, 5 and 11 MeV at high densities $\rho Y_e$ = 10$^{8}$, 10$^{9}$ and 10$^{10}$ g/cm$^{3}$, respectively,
decreasing slightly as the temperature increases.
The reaction rates become larger at higher densities because of the larger
chemical potential.

\begin{figure}
\begin{center}
\includegraphics[width=9.5cm]{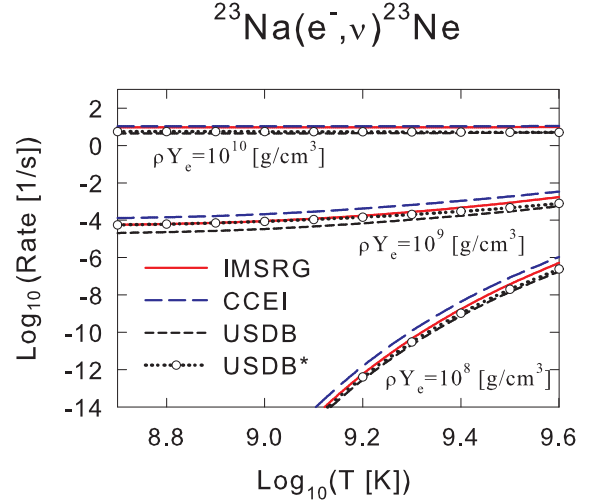} 
\end{center}
\caption{
Calculated electron capture rates on $^{23}$Na obtained by shell model calculations with different effective interactions.}
\label{Ne23ec}
\end{figure}

 Here, we evaluate the electron capture rates on $^{23}$Na and $^{25}$Mg.
These rates are important in the study of the nuclear URCA processes that determine the cooling of the O-Ne-Mg core
of stars with initial masses of 8-10 M$_{\odot}$ \cite{Toki2013,STN2016}.
The electron-capture rates for $^{23}$Na(e$^-$,$\nu$)$^{23}$Ne are evaluated by using the $B(GT)$ strengths obtained from the
IM-SRG and CCEI methods for densities 
$\rho Y_e$ =$10^8$, $10^9$ and $10^{10}$ g/cm$^3$ and temperatures T = 10$^{8.7}$-10$^{9.6}$ K.
 The GT transitions from 3/2$^+$(g.s.) and 5/2$^+$(0.440 MeV) states in $^{23}$Na are included.
The calculated rates are shown in Fig. \ref{Ne23ec}. Here the same quenching factor $f_q = 0.77$ is used
for all three interactions.
The rates calculated by using the IM-SRG and CCEI methods are large compared with the USDB results.  
 In the USDB* interaction  shown in Fig. \ref{Ne23ec}, the available experimental energies and $B(GT)$ strengths are taken into account, so the USDB* is more realistic  \cite{STN2016}.
Our results for the IM-SRG are close to those for the USDB*.
This comes from the fact that the $B(GT)$ value for the transition from the g.s. of $^{23}$Na to the g.s. of $^{23}$Ne is close to the experimental value in case of IM-SRG, while it is smaller (larger) in case of USDB (CCEI).
Both IM-SRG and CCEI give larger $B(GT)$ than USDB and the experiment for $E_x$ = 0.5-3.5 MeV. 
Compared to USDB and USDB*, this leads to an enhancement of the capture rates by about a factor 2 at higher densities, 
$\rho Y_e$ =10$^{10}$ g/cm$^{3}$.
Since the dominant contribution to the capture rates for $^{23}$Na ($e^{-}$, $\nu$) $^{23}$Ne comes from the g.s. to g.s. transition \cite{Toki2013,STN2016}, IM-SRG is practically applicable to the evaluation of the weak rates in stellar environment, in spite of the enhanced $B(GT)$ strength at $E_x$ = 0.5-3.5 MeV.    
The results calculated from the CCEI are enhanced compared to the USDB* by a factor of 2-4.

\begin{figure}
\begin{center}
\includegraphics[width=9.5cm]{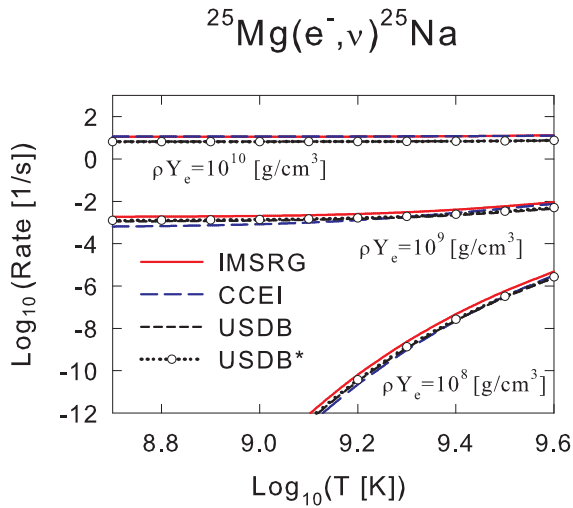} 
\end{center}
\caption{
Calculated electron capture rates on $^{25}$Mg obtained by shell model calculations with different effective interactions.}
\label{Na25ec}
\end{figure}

Electron-capture rates for $^{25}$Mg(e$^-$,$\nu$)$^{25}$Na  are shown in Fig. \ref{Na25ec}.
The GT transitions from 5/2$^+$(g.s.), 1/2$^+$(0.588 MeV) and 3/2$^+$(0.975 MeV) states in $^{25}$Mg are taken into account. 
The rates calculated with the CCEI and IM-SRG are close to those of the USDB* within a factor of 2.
We thus find that the GT strengths obtained by the $ab~initio$ interactions are reasonably valid for the evaluation
of the weak rates at high densities and high temperatures for the lower mass $sd$ shell nuclei considered here.

\newpage

\section{Conclusions }

In the present work we used $ab~initio$ effective interactions to calculate the GT strengths in the $sd$ shell nuclei. 
The results of the USDB interaction show reasonable agreement with the available experimental 
data in comparison to $ab~initio$ effective interactions.
Our work adds more information on the GT strength distributions obtained in earlier work. In some cases shifting of
energy levels occurs,
because $ab~initio$ effective interactions are not able to reproduce correctly the excited states at the particular 
observed energies.

 The GT calculated strengths are found to be applicable to evaluate nuclear weak rates for some lower mass $sd$ shell 
nuclei, such as $^{23}$Na and $^{25}$Mg, 
 within a factor of 2-4 in stellar environments. These nuclear weak rates play important 
 roles in astrophysical processes.
 It is highly desirable to improve the $ab~initio$ method by including further 
the three-body valence cluster terms, that is, the terms IM-SRG(3) or $H_3^{A_c+3}$.
 It is also of interest to extend the method to include contributions from the two-body GT operators. In this work, 
 they are taken into account by a phenomenological quenching factor for the one-body operator.

\section*{Acknowledgments}
The authors would like to thank W. Bentz for the careful reading of the manuscript.
We would like to thank P. Navr\'atil, S. R. Stroberg and G. R. Jansen  for useful discussions on $ab~initio$ effective interactions.
AS acknowledges financial support from MHRD for her Ph.D. thesis work. PCS would like to thank financial support from faculty
initiation grant.
 TS would like to thank a support from the Grants-in-Aid for Scientific Research (15K05090) of the MEXT of Japan.

\end{document}